\documentclass[aps,preprint,showpacs,preprintnumbers,amsmath,amssymb,showkeys,
  floatfix]{revtex4}

\usepackage{dcolumn}   % needed for some tables
\usepackage{bm}        % for math
\usepackage{amssymb}   % for math
\usepackage{amsmath}
\usepackage{dsfont}
\usepackage{graphicx,psfrag,amsmath,subfigure}
\usepackage{color}
\usepackage{float}

\definecolor{webgreen}{rgb}{0,.5,0}
\definecolor{webbrown}{rgb}{.6,0,0}
\usepackage{hyperref}
\hypersetup{colorlinks=true,
            breaklinks=true,
            plainpages=false, 
            bookmarksnumbered, 
            bookmarksopen=true,
            bookmarksopenlevel=1,
            urlcolor=webbrown, 
            linkcolor=blue, 
            citecolor=webgreen
            }

\newcommand{\be}{\begin{equation}}
\newcommand{\ee}{\end{equation}}
\newcommand{\bea}{\begin{eqnarray}}
\newcommand{\eea}{\end{eqnarray}}
\newcommand{\bee}{\begin{eqnarray*}}
\newcommand{\eee}{\end{eqnarray*}}
\def\dd{{\rm d}}
\newcommand\half{\textstyle\frac{1}{2}}
\renewcommand\Re{{\rm Re}}

\begin{document}

\title{Measuring the trilinear neutral Higgs boson couplings in the 
minimal supersymmetric standard model at $e^+ e^-$  colliders 
in the light of the discovery of a Higgs boson}

\author{  Charanjit K. Khosa and P. N. Pandita}
\affiliation{Centre for High Energy Physics, Indian Institute of
Science, Bangalore 560 012, India}

\date{\today}

\begin{abstract}
We consider the measurement of the trilinear couplings
 of the neutral Higgs bosons in the Minimal Supersymmetric
 Standard Model (MSSM) at a high energy  $e^+ e^-$ linear 
collider in the light of the discovery of a Higgs
 boson at the CERN Large Hadron Collider~(LHC). We identify
 the state observed at the LHC with the lightest
 Higgs boson  ($h^0$) of the MSSM, and impose the constraints following from this
 identification, as well as other experimental constraints on the MSSM
 parameter space. In order to measure trilinear
 neutral Higgs couplings, we consider different processes where the heavier
 Higgs boson ($H^0$) of the MSSM is produced in electron-positron
 collisions, which subsequently decays into a pair of lighter Higgs boson. We identify
 the regions of the MSSM parameter space where it may be possible to measure
 the trilinear couplings of the Higgs boson at
 a future electron positron collider. A measurement of the trilinear
Higgs couplings	is a crucial step in the construction of the Higgs potential,
and hence  in establishing the phenomena of spontaneous symmetry breaking
in gauge theories.
\end{abstract}

 {}
 \maketitle
 \newpage
\section{ Introduction}

The ATLAS \cite{Aad:2012tfa} and CMS \cite{Chatrchyan:2012ufa}
experiments at the CERN Large Hadron Collider
have independently observed a resonance at about 125-126 GeV. The
 ATLAS experiment after collecting data at an integrated
luminosity of 4.8 fb$^{-1}$ at $\sqrt{s}$ = 7 TeV, and 5.8
fb$^{-1}$ at $\sqrt{s}$ = 8 TeV confirmed the evidence for
the production of a neutral boson with a measured mass of 126.0
$\pm$ 0.4(stat) $\pm$ 0.4(syst) GeV, with a significance of 5.9
$\sigma$. The CMS experiment after collecting 5.1 fb$^{-1}$  at 7
TeV, and 5.3 fb$^{-1}$ at 8 TeV reported an evidence of a neutral
boson at 125.3 $\pm$ 0.4(stat) $\pm$ 0.5(syst) GeV, with a
significance of 5.8 $\sigma$. The CMS and ATLAS collaborations
have now combined their analysis, 
which leads to the mass of the reported resonance
as~\cite{atlascms}
\bea
m_h & = & 125.09 \pm  0.21 \mbox{(stat)} \pm 0.11 \mbox{(syst)} \,\,\mbox{GeV}.
\eea 
The properties of the discovered state are consistent with the properties 
of the  standard model (SM) Higgs boson. Nevertheless, this discovery opens
up the possibility of searches for new physics beyond the standard model.
Since the  Higgs sector suffers from the problems of naturalness 
and hierarchy, a light Higgs boson is technically unnatural in standard model. 
Supersymmetry~(SUSY)~\cite{Wessbagger} is
at present the leading candidate for physics beyond the SM  in which
a light Higgs boson, as discovered at the CERN LHC, is technically
natural. The simplest implementation~\cite{Nillesnath,Nath:1983fp}
of the idea of supersymmetry
at weak scale is the minimal supersymmetric standard model~(MSSM).
In MSSM the Higgs sector is more complex than  the 
SM Higgs sector, and consists of two Higgs doublet 
superfields~($H_1$ and $H_2$). After gauge symmetry breaking 
the physical Higgs sector consists of two $CP$-even Higgs bosons~($h^0, H^0;
m_{h^0} < m_{H^0}$), one $CP$-odd Higgs boson~($A^0$), and two charged 
states~($H^\pm$). Although gauge invariance and supersymmetry fix the quartic
couplings of the Higgs bosons in the MSSM in terms of the $SU(2)_L$ and 
$U(1)_Y$ gauge couplings, $g$ and $g'$, respectively, there remain
two independent parameters that describe the Higgs sector of the MSSM.
These are conveniently chosen to be  the mass of  
the $CP$-odd Higgs boson ($m_{A^0}$), and the ratio of the vacuum expectation
values of the neutral components of the two Higgs fields, 
$\tan\beta  \equiv  <H_2^0>/<H_1^0>$. All the Higgs boson masses and the Higgs
couplings in the MSSM can, then, be described, at the tree level, 
in terms of these two parameters.  Discovery of more than one Higgs
boson will, thus,  be an indication of an extension of the SM, SUSY
 being the most favored candidate. It is, therefore, important to
try to discover an extended Higgs sector, if it exists,  at the LHC.
On the other hand it is also crucial to discover the supersymmetric partners 
of the SM particles, including the squarks, gluinos and sleptons,
as well as the neutralinos and charginos.  

In the absence of any signal for supersymmetric partners of the SM particles,
it is appropriate to ask the question whether nonobservation of an 
extended Higgs sector would imply that the new physics is at a much higher 
energy scale. This question is related to  the decay 
patterns~\cite {PanditaPatra,AnanthLahiriPandita,AnanthLahiriPanditaPatra} 
of the Higgs bosons of MSSM.
This in turn is dependent on the Higgs couplings in the MSSM, and their 
deviation from the corresponding SM Higgs couplings. Apart from the 
couplings of Higgs bosons to the SM particles, which can be different
from those of the corresponding SM Higgs couplings, the trilinear
Higgs couplings can be very different in MSSM as compared to the 
SM trilinear Higgs coupling.

The measurement of the trilinear Higgs 
couplings~\cite{dhz,OPtricooup,Osland:1999ae,Osland:1999qw,Zerwas,Wu:2015nba} is an important
step in the reconstruction of the Higgs potential, and thereby
confirm the Higgs mechanism as the origin of the spontaneous
breaking of the gauge symmetries in the SM~(and in MSSM). As 
explained above, the tree level Higgs sector of the MSSM is described
by only two parameters~($m_{A^0}$ and $\tan\beta$). 
At the tree level the trilinear Higgs self couplings of the MSSM can
also  be described in terms of the same two parameters. In this paper we consider the question of 
measurement of the trilinear Higgs couplings in the MSSM 
at a high energy electron-positron collider in the light of the
observation of the Higgs boson at the CERN LHC, which we identify with the 
lightest Higgs boson of the MSSM.

The plan of this paper is as follows. In Section \ref{hsmssm} we recall
 the basic features of the Higgs sector of the minimal supersymmetric
 standard model  and discuss the implications of the 125 GeV Higgs
 identification for the mass of the heavier
 Higgs boson ($H^0$) of the model, especially its dependence on the
 parameter space of the MSSM. In Section \ref{thcparaspace} we summarize
 the tree level as well as radiatively corrected trilinear Higgs couplings
 in the MSSM and discuss their dependence on the different
 parameters. In Section \ref{higgprodanalysis}, we estimate the
 cross-section for different heavy Higgs production processes and discuss its
 branching ratios to light Higgs pair in the MSSM parameter space. In
 Section \ref{tricoupmeasurement} we
 highlight the regions in the ($m_{A^0},\tan\beta$) plane
 where some of the MSSM Higgs trilinear couplings can be measured.
 In Section \ref{Conclusions} we summarize our results and conclusions.

\section{The Higgs sector of the minimal supersymmetric standard model\label{hsmssm}}
To begin with, and to have a proper perspective,
we recall that the potential for the physical
Higgs boson in the SM can be written as

\be  V^h_{SM}=\lambda \left(\phi^2 -\frac{v^2}{2}\right)^2=\frac{m_h^2}{2}
 h^2 +
  \lambda_{hhh}^{SM} \frac{h^3}{3!}+\lambda_{hhhh}^{SM}  \frac{h^4}{4!},
  \label{higgspotential} \ee
\noindent
where $<\phi> = v/\sqrt 2$ $\approx$ 174 GeV is the vacuum expectation value of the neutral 
component of the Higgs doublet $\left(\phi=(1/\sqrt{2}) 
(v+h)^T\right)$, and $m_h$ is the mass of the physical Higgs boson. 
The trilinear and quartic Higgs
couplings in the standard model can, then, be written as
(in units of $ (\sqrt{2} G_F)^{1/2} m_Z^2 $ = 33.77 GeV and
 $(\sqrt{2} G_F)m_Z^4$ = 1140.52 GeV$^2$, respectively)  
\begin{eqnarray} 
\lambda_{hhh}^{SM} & = & \frac{3 m_h^2}{m_Z^2}=5.6454, \\
\lambda_{hhhh}^{SM} & = & \frac{3 m_h^2}{m_Z^4}=0.00068\,\, \mbox{GeV}^{-2}. \label{SMtandqcoup} 
\end{eqnarray}
The Higgs sector of the minimal supersymmetric model 
consists of two Higgs superfields, $H_1$ and $H_2$, with opposite
 hypercharge ($Y$ = -1, $Y$ = 1, respectively). The mass matrix for the $CP$ even Higgs bosons 
can be written as~\cite{Ellis:1991zd,Okada:1990vk,Haber:1990aw}
\begin{eqnarray}
\label{Eq:M2} {\mathcal M}^2 &=&  \left[ \begin{array}{cc} m_{A^0}^2
\sin^2 \beta + m_Z^2 \cos^2\beta &
-(m_Z^2 + m_{A^0}^2) \sin\beta \cos \beta\\
-(m_Z^2 + m_{A^0}^2) \sin\beta \cos \beta & m_{A^0}^2 \cos^2 \beta + m_Z^2
\sin^2\beta
\end{array} \right] \nonumber \\
& & + \frac{3 g^2}{16 \pi^{2} m_W^2} \left[ \begin{array}{cc}
\Delta_{11} & \Delta_{12}\\
\Delta_{12} & \Delta_{22}
\end{array} \right],
\label{massmatrix1}
\end{eqnarray}
where the second matrix embodies the radiative corrections.
The radiative corrections $\Delta_{ij}$ depend, 
besides the top- and bottom-quark and squark masses, on the Higgs(ino) bilinear
parameter $\mu$ in the superpotential, the soft supersymmetry breaking 
trilinear couplings~($A_t, A_b$), and soft supersymmetry
breaking scalar masses~($m_Q, m_U, m_D$), as well as on $\tan\beta.$
The radiatively corrected $CP$-even masses  are
 obtained by diagonalizing the 
$2 \times 2$ mass matrix (\ref{massmatrix1}). After diagonalization the
masses and the mixing angle $\alpha$ in the $CP$ even Higgs sector 
can be written as
\begin{eqnarray} m_{h^0/H^0}^2&=&
\frac{1}{2}(m_{A^0}^2+m_Z^2+\Delta M_{11}^2+ \Delta M_{22}^2 \mp
\sqrt{m_{A^0}^4+m_Z^4-2 m_{A^0}^2 m_Z^2 \cos 4 \beta + \delta}),
\nonumber \\ \nonumber 
\tan\alpha&=&\frac{2 \Delta M_{12}^2-(m_{A^0}^2+m_Z^2)\sin
2\beta}{\Delta M_{11}^2-\Delta M_{22}^2+(m_Z^2-m_{A^0}^2)\cos 2
\beta+\sqrt{m_{A^0}^4+m_Z^4-2 m_{A^0}^2 m_Z^2 \cos 4 \beta+\delta}},
\\ \vspace{1.5cm}\nonumber
\delta&=&4\Delta M_{12}^4+(\Delta M_{11}^2-\Delta
M_{22}^2)^2-2(m_{A^0}^2-m_Z^2)(\Delta M_{11}^2-\Delta M_{22}^2)\cos
2\beta\\\nonumber &&-4(m_{A^0}^2+m_Z^2)\Delta M_{12}^2\sin 2\beta,
\\  &&
 \Delta M_{ij}^2=\frac{3 g^2}{16 \pi^{2} m_W^2} \Delta_{ij}, 
\qquad i,j=1,2. \label{higgsmassndangle}
\end{eqnarray}
\noindent
In the numerical
 calculations of the Higgs boson mass, we shall take into account the one-loop
 and leading two-loop radiative corrections \cite{Ellis:1991zd,Okada:1990vk,Haber:1990aw,Carena:1995bx,Haber:1996fp,Carena:2000dp}. 
We shall identify the lightest $CP$-even Higgs boson of the MSSM with 
the state discovered at the CERN
LHC~\cite{Aad:2012tfa,Chatrchyan:2012ufa,atlascms} at $m_{h^0}$ $\approx$ 125 GeV. 
Requiring that the production cross section of the $125$ GeV Higgs boson
decaying to two photons agrees with the one observed at the CERN LHC divides
the MSSM Higgs parameter space into two distinct regimes~\cite{Li:2013nma}:
(i) The nondecoupling regime where $m_{A^0}\le 130$ GeV. In this region
the heavier Higgs $H^0$ is SM-like, and the light Higgs $h^0$ and $CP$-odd
Higgs $A^0$ are almost degenerate \cite{Pandita:1984hf}, with mass close to 
$m_Z,$ while the charged Higgs bosons $H^{\pm}$ are nearly 
degenerate with $H^0.$ 
For a discussion of this possibility and its phenomenological 
implications at the LHC, 
see e.g.~\cite{Boos:2002ze,Boos:2003jt,Christensen:2012ei}.
(ii) The decoupling regime where $m_{A^0} > 300$ GeV. In this limit
the light $CP$-even Higgs boson $h^0$ is SM-like, and all other physical
Higgs bosons are heavy, and almost degenerate~\cite{Haber:1995jm} 
with $A^0.$ These decoupling properties  are a generic
 feature of models with extended Higgs sectors.

The nondecoupling scenario has also been explored recently in
 context of LHC 8 TeV data \cite{Bhattacherjee:2015qct}. But as far as measurement of  
MSSM Higgs bosons trilinear coupling is concerned, the decoupling scenario is a 
more viable scenario. Therefore, in the following we shall consider only the
MSSM with decoupling scenario while discussing the trilinear Higgs couplings.
The nondecoupling scenario is discussed in some detail 
in~\cite{PanditaPatra}.
 
\begin{figure}
\subfigure[]{
\includegraphics[scale=.40]{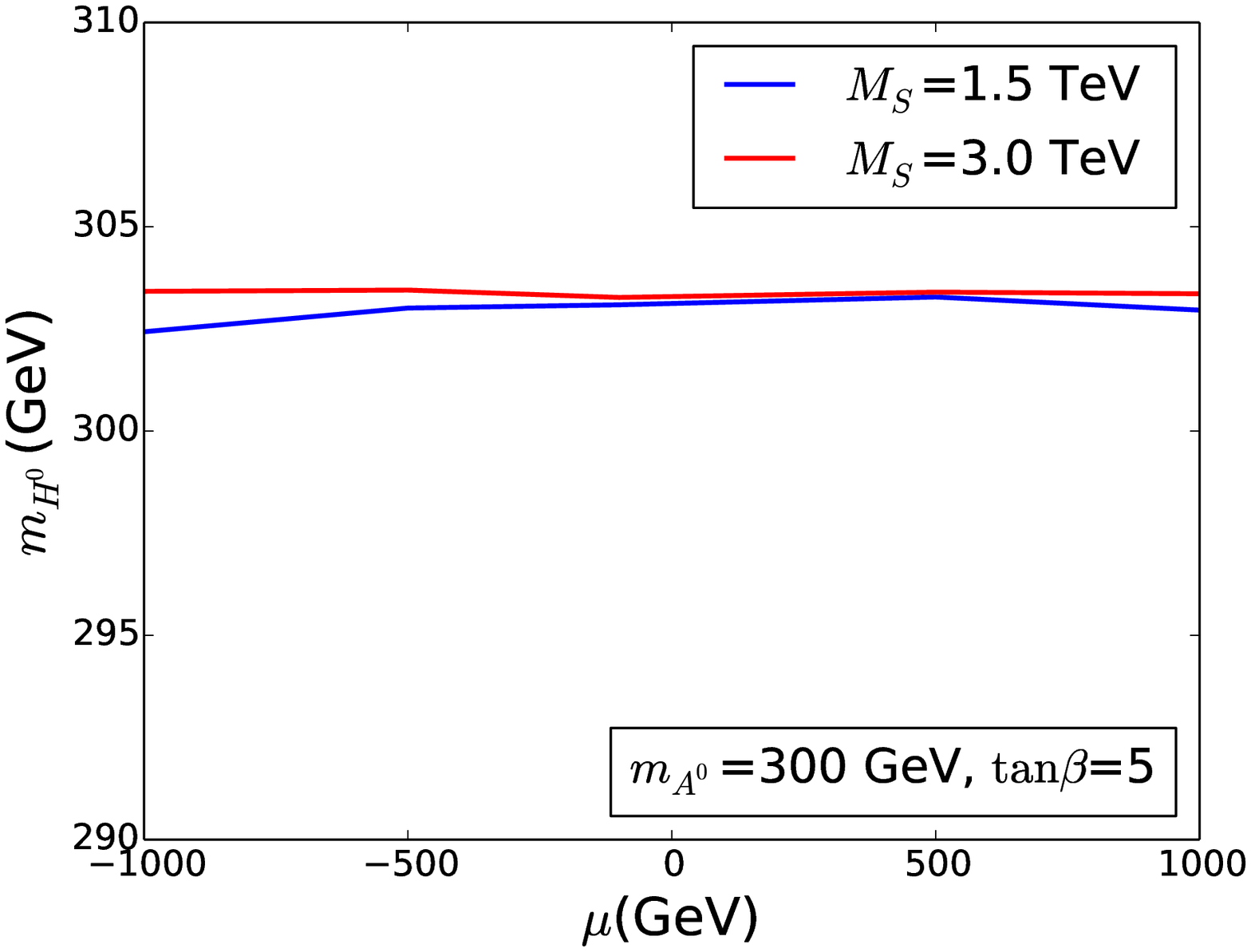}}
\hspace{-.7cm}\subfigure[]{
\includegraphics[scale=.40]{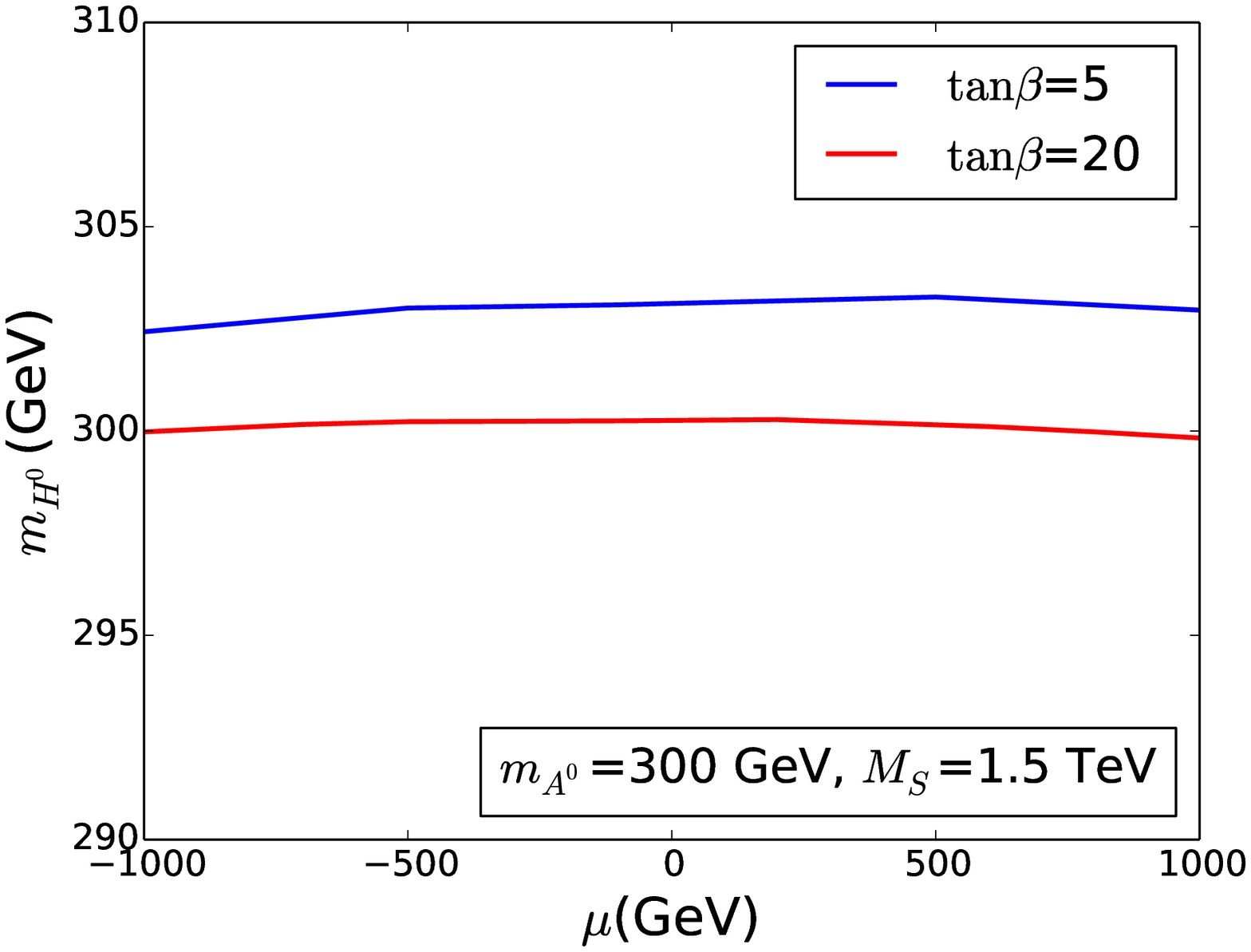}}
\subfigure[]{
\includegraphics[scale=.38]{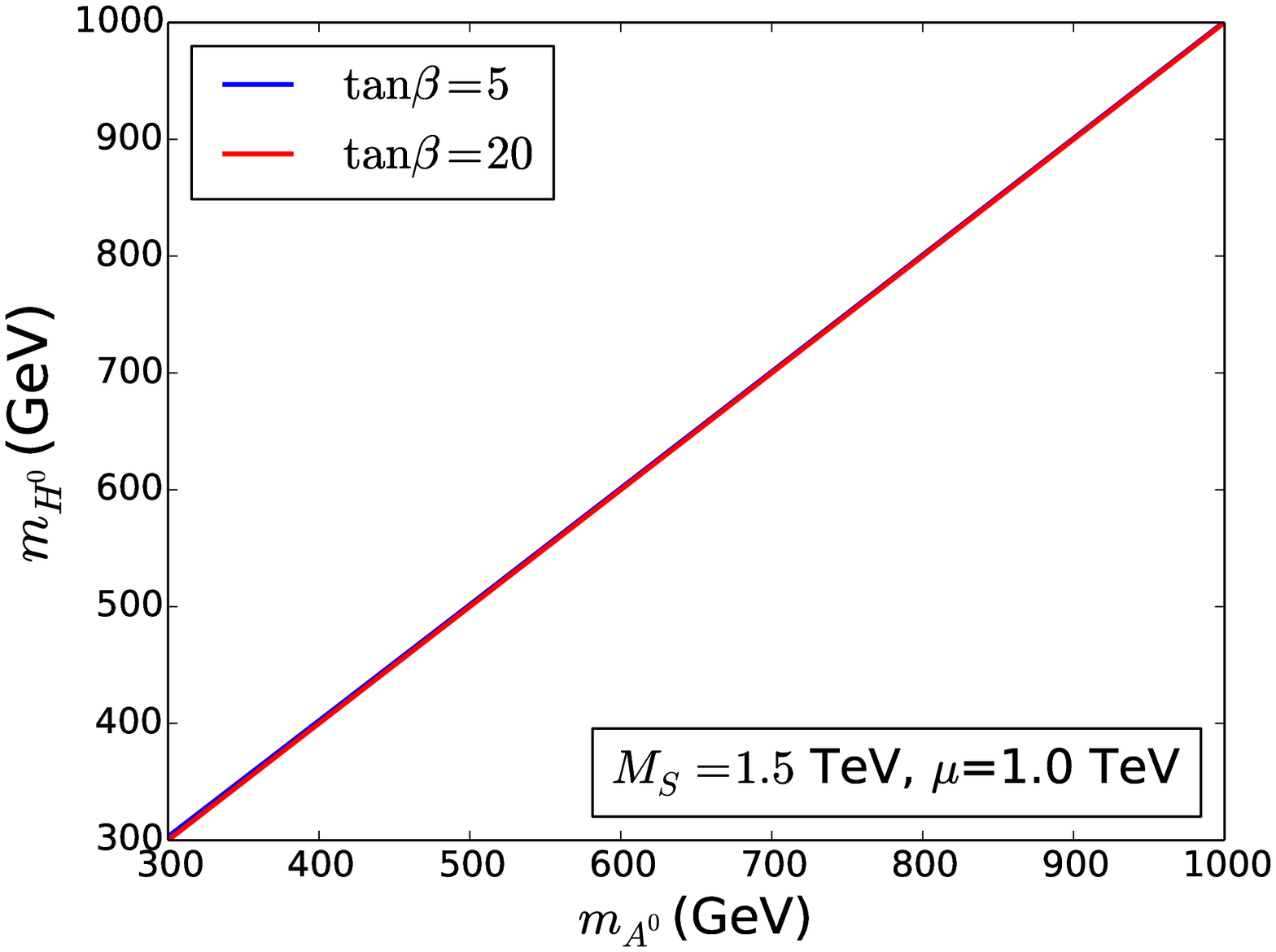}}
\caption{The mass of heavier Higgs boson $m_{H^0}$ as a
 function of (a) $\mu$ for fixed $\tan\beta$;
(b) as a function of  $\mu$ for fixed $M_S$; and (c) as a function of 
$m_{A^0}$ for the values of the other
parameters as shown in the inset. Soft trilinear parameter ($A_t$)
is adjusted to get lightest Higgs boson ($h^0$) mass in 122-128 GeV
range.}\label{fig1}
\end{figure}

\begin{table}
 $$
 \begin{array}{|c|c|c|}
\hline
\mu & A_t & m_{h^0} \\
\mbox{in GeV} & \mbox{in GeV} & \mbox{in GeV} \\
\hline
-1000  & 3845 & 123.25 \\
-500 & 3845 &   123.61 \\
500 & 3545 & 123.86 \\
1000 & 3545 & 123.09 \\
\hline
\end{array}
 $$
\caption{Values of $\mu$, $A_t$ and Higgs mass for $M_S$=1.5 TeV, $m_{A^0}$=300 GeV and $\tan\beta$=5.}
 \label{tableat} \end{table}

In order to  identify the state observed at the CERN
LHC with the lightest Higgs boson of the MSSM, we shall 
adjust the  supersymmetry breaking soft trilinear parameter $A_t$ so as to 
get $m_{h^0}$ in the range 122-128 GeV, corresponding to a 3 GeV
 theoretical uncertainty in the Higgs mass calculations. 
Having fixed $m_{h^0}$, we have performed 
a numerical scan of the MSSM parameter space using CalcHEP \cite{Belyaev} which
 uses SuSpect \cite{Djouadi:2002ze} for MSSM spectrum calculations. SuSpect checks the
 stability of potential and calculates the spectrum only for stable points. 
For our analysis we use set of input parameters which are consistent
 with known experimental constraints,
and also which have the possibility of leading to the supersymmetry spectra 
that may be observable in the upcoming experiments. We do this in order to have 
low energy supersymmetry as a viable  option for solving the naturalness 
and hierarchy problem of the standard model. Using this procedure,
we have calculated the dependence of the heavy $CP$-even Higgs boson
mass $m_{H^0}$ on $\mu$ for different 
values of $\tan\beta,$ and the SUSY breaking scale $M_S$, which
 is defined to be $\sqrt{m_{\tilde{t_1}}m_{\tilde{t_2}}}$, where
 $m_{\tilde{t_1}}$ and  $m_{\tilde{t_2}}$ are the masses of the two stop states.
This dependence is shown in Fig.\ref{fig1}. This Fig. shows that $m_{H^0}$ does
 not depend significantly on $\mu$. For a 
given value of $\tan\beta$ it depends weakly on $M_S$. However, $m_{H^0}$
has a significant dependence on $m_{A^0}$ and $\tan\beta$, and can be 
described in terms of these two parameters to a good accuracy
 when we use the fact that $m_{h^0}$ 
lies in the range 122-128 GeV. As an example, for one
 set of $M_S$, $\tan\beta$ and $m_{A^0}$ we show the values of the parameter $A_t$ 
 with different values of $\mu$ in Table \ref{tableat}. For the considered range of
 $\mu$, $A_t$ does not change much. For $M_S$=1.5 TeV, typical
 range of $A_t$ is 2500-4000 GeV depending on $\tan\beta$.

\section{Trilinear Higgs Boson Couplings in the Minimal
 Supersymmetric Standard Model\label{thcparaspace}}
We now discuss the question of the measurement of trilinear couplings of
 the neutral Higgs bosons ($h^0,H^0$) of the minimal
 supersymmetric standard model. These couplings
 receive contributions at the tree
level as well as from radiative corrections. We shall assume 
CP conservation throughout in this paper. Then the trilinear 
couplings can be written as\cite{Barger:1991ed}
\begin{eqnarray}
\lambda_{hhh} & = & \lambda_{hhh}^0 + \Delta \lambda_{hhh},
 \nonumber \\
\lambda_{Hhh} & = & \lambda_{Hhh}^0 + \Delta \lambda_{Hhh},
 \nonumber\\
\lambda_{hAA} & = & \lambda_{hAA}^0 + \Delta\lambda_{hAA},
 \nonumber \\
\lambda_{HAA} & = & \lambda_{HAA}^0 +
\Delta\lambda_{HAA},  \nonumber \\
 \lambda_{HHH} & = & \lambda_{HHH}^0 +
\Delta\lambda_{HHH},  \nonumber \\
\lambda_{HHh} & = & \lambda_{HHh}^0 + \Delta\lambda_{HHh},
\label{thcmssmtl}
\end{eqnarray}
where $\lambda^0$'s are the tree level values of the couplings, and $\Delta\lambda$'s are
 the radiative corrections. In this paper we shall consider the
 one-loop approximation for the radiative corrections to the trilinear
 Higgs couplings. The leading two loop SUSY-QCD  corrections to the trilinear 
couplings are available in the literature \cite{Brucherseifer:2013qva}
which reduces the scale dependence of one-loop corrections but the 
contribution is very small as compared to the one-loop corrections. The
 tree-level couplings in units of $(\sqrt{2}G_F)^{1/2}m_Z^2$
can be written as 
\begin{eqnarray}
\lambda_{hhh}^0 & = & 3 \cos2\alpha \sin(\beta + \alpha), \nonumber \\
\lambda_{Hhh}^0 & = & 2\sin2\alpha \sin(\beta + \alpha) - \cos
2\alpha
\cos(\beta + \alpha), \nonumber \\
 \lambda_{hAA}^0 & = & \cos2\beta \sin(\beta
+ \alpha), \label{Eq:lambda-hAA0} \nonumber \\
\lambda_{HAA}^0 & = & -\cos2\beta \cos(\beta + \alpha),
 \nonumber \\
\lambda_{HHH}^0 & = & 3\cos2\alpha \cos(\beta + \alpha),
\nonumber \\
\lambda_{HHh}^0 & = & -2\sin 2\alpha \cos(\beta +
\alpha)-\cos2\alpha \sin(\beta + \alpha),
\label{thcmssmfull}
\end{eqnarray}

\begin{figure}
\subfigure[]{
\includegraphics[scale=0.41]{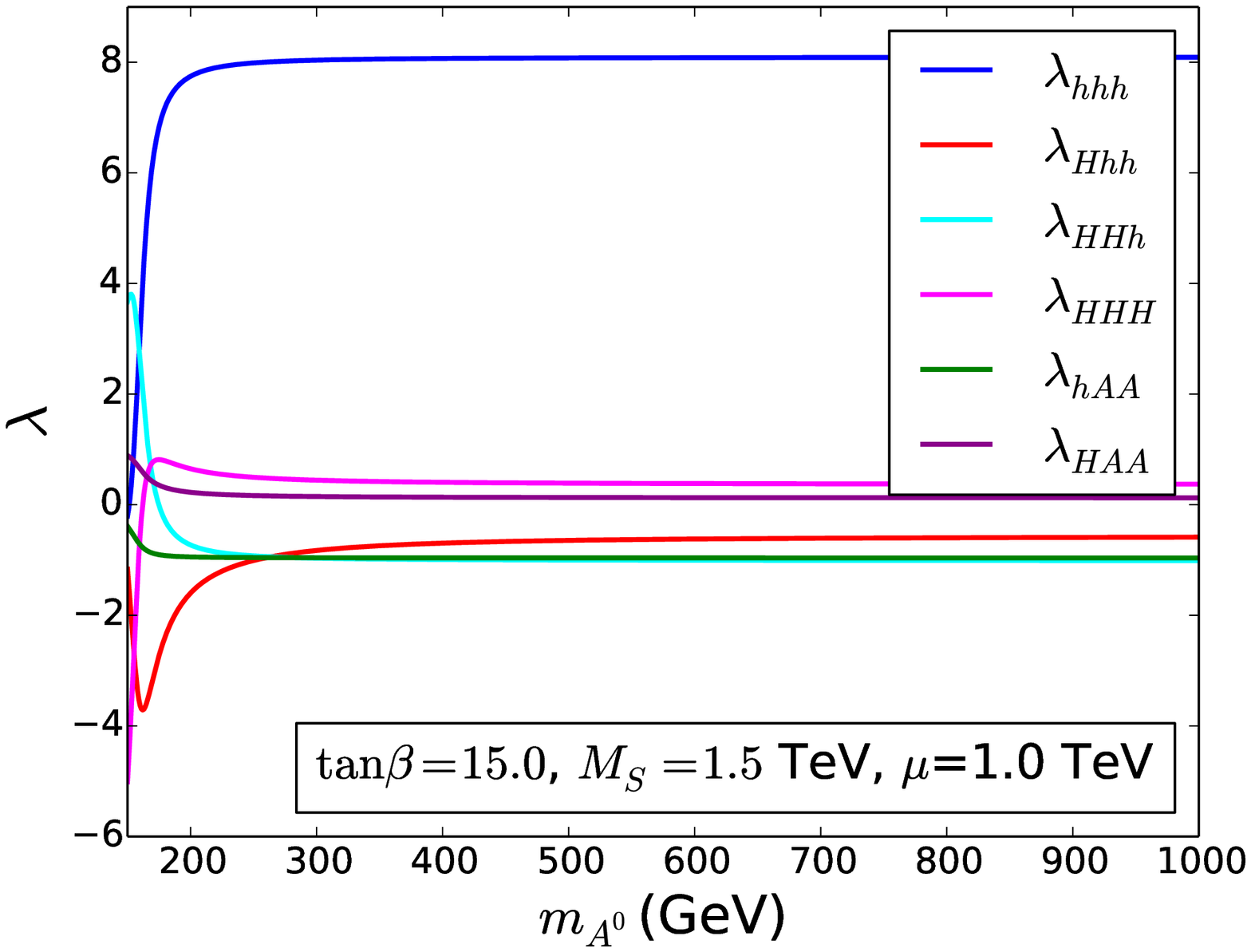}}
\hspace{-0.6cm}\subfigure[]{
\includegraphics[scale=0.41]{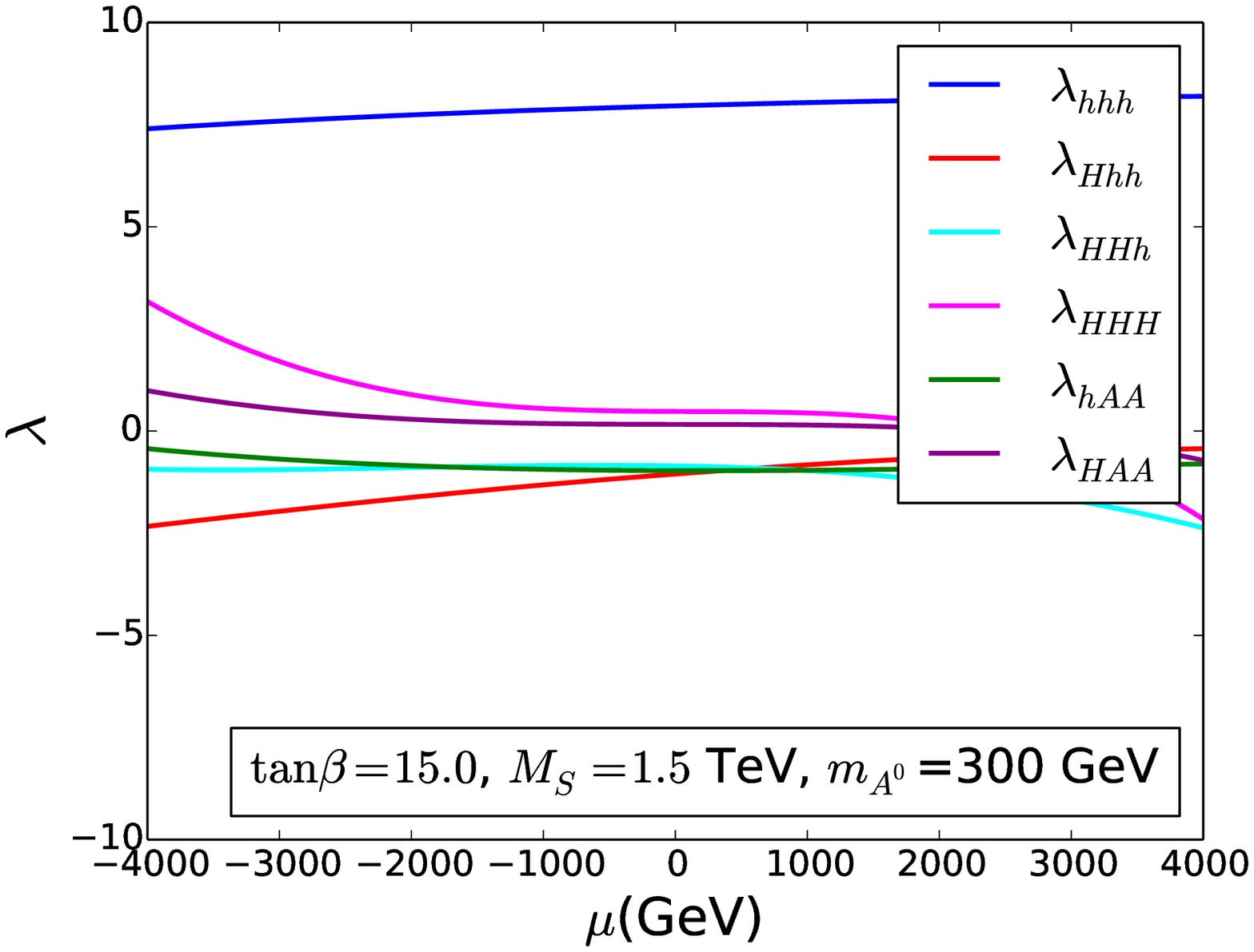}}
\subfigure[]{
\includegraphics[scale=0.41]{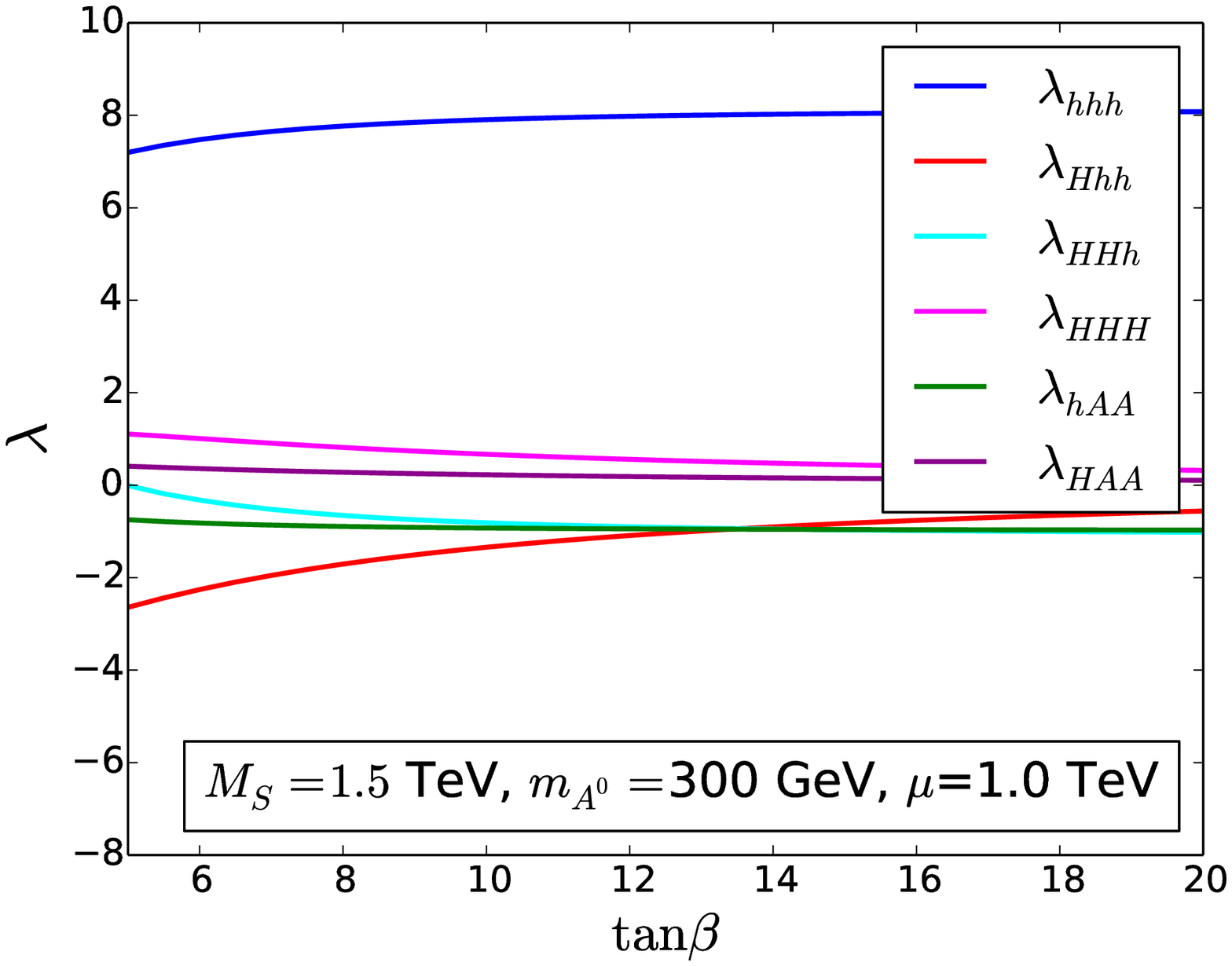}}
 \caption{Variation of radiatively corrected trilinear couplings in MSSM
 as functions of (a) $m_{A^0}$;  (b) $\mu$ ; and (c) $\tan\beta$ for the
 values of the parameters as shown in the inset. In all
  the figures the soft trilinear parameter ($A_t$) is adjusted to
  obtain the value of the lightest Higgs boson
 mass in 122-128 GeV range. In this Fig. and in the following we
 have taken $M_S$ =1.5 TeV.}\label{tcoupling1}
\end{figure}
\noindent
where $\alpha$ is the mixing angle in the CP-even Higgs sector, which can be
obtained from the diagonalization of mass matrix (\ref{massmatrix1}), 
as shown in (\ref{higgsmassndangle}). The radiative corrections $\Delta\lambda$'s
in (\ref{thcmssmtl}) are summarized in the Appendix.

\begin{figure}
\centering
\includegraphics[scale=0.5]{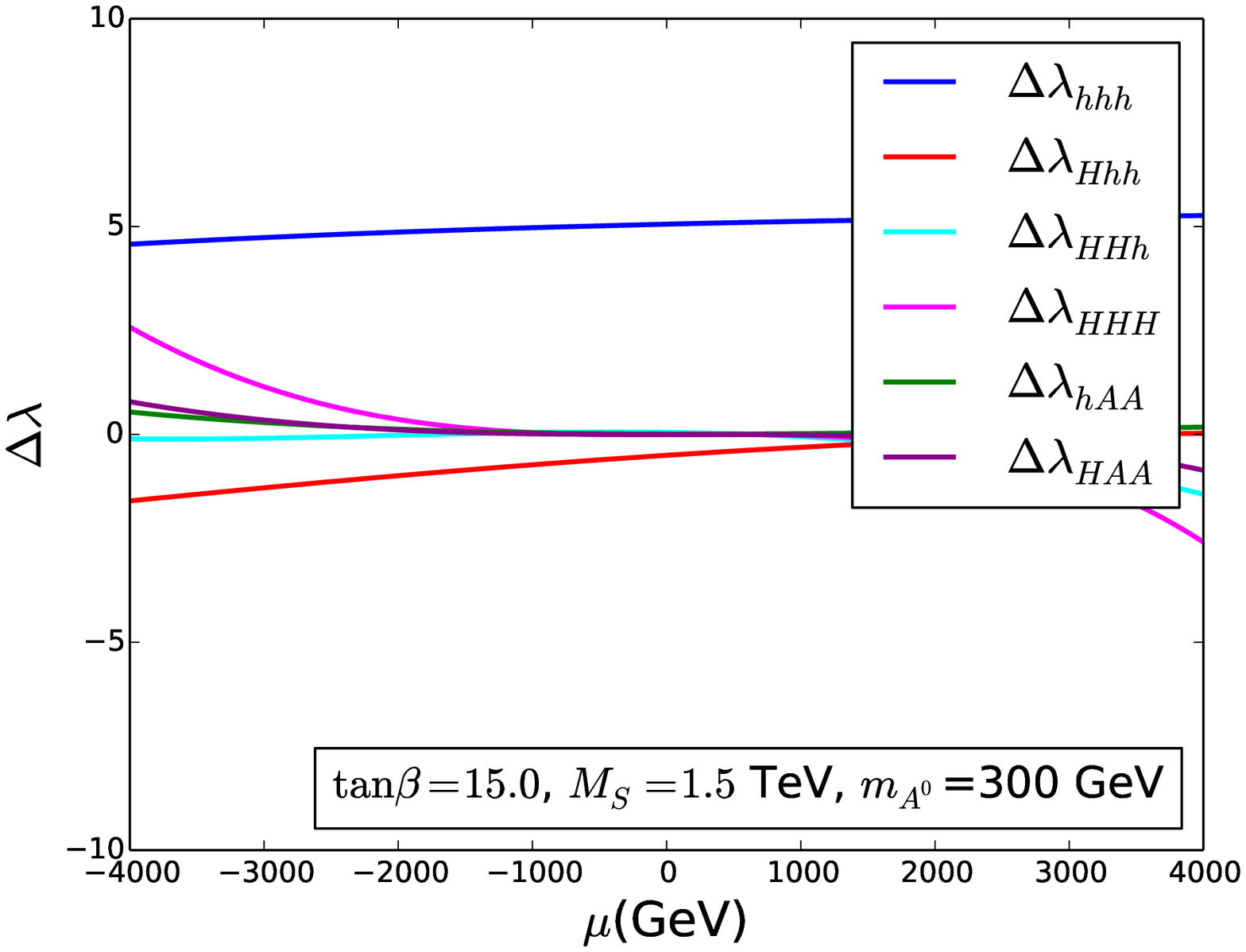}
\caption{Variation of the radiative corrections to the trilinear
Higgs couplings ($\Delta \lambda$'s), in units of
 $(\sqrt{2} G_F)^{1/2}m_Z^2$, in MSSM as a function of
$\mu$ for the fixed values of the other parameters as shown in the inset.
The value of $A_t$ is adjusted to get lightest Higgs boson mass in 122-128 GeV
range.}\label{tcoupling2}
\end{figure}

We note that the Higgs sector depends on five parameters in the
MSSM, $m_{A^0}$ and $\tan\beta$ from tree level mass matrix, and three 
parameters $A_t$, $\mu$ and $M_S$ from radiative corrections. 
Of these $M_S$ is fixed from
nonobservation of colored particles to be greater
 than 1.5 TeV, and $A_t$ is used to fix the value of lightest 
Higgs mass ($m_{h^0}$). For fixed value of
$\tan\beta$ we are left with two parameters $m_{A^0}$ and $\mu$.
In Fig. \ref{tcoupling1} we show the variation of radiatively corrected trilinear
 couplings with
respect to $m_{A^0}$ (for fixed value of $\mu$), and with respect to $\mu$
(for fixed value of $m_{A^0}$), respectively. In Fig. \ref{tcoupling1} (c)
 we show the trilinear couplings
 as a functions of $\tan\beta$ (for fixed value of $\mu$ and $m_{A^0}$).
Most of the
variation in the trilinear Higgs couplings comes from the
variation of the radiative corrections, a fact which is shown
 in Fig. \ref{tcoupling2}, where we plot only the
radiative corrections to different trilinear Higgs couplings.

From these Figures, we can see that the trilinear couplings are sensitive to
$m_{A^0}$ upto 500 GeV except the ones involving CP-odd
 Higgs boson. In Fig. \ref{tcoupling1} (c), we have shown
 the variation of trilinear Higgs couplings as a function of
 $\tan\beta$ for a value of $m_{A^0}$ = 300 GeV, with other
 parameters kept fixed, and  this plot shows that 
trilinear couplings are weakly dependent on $\tan\beta$. In this paper
 we shall consider only the trilinear couplings $\lambda_{hhh}$ and
 $\lambda_{Hhh}$ between the neutral Higgs bosons $h^0$ and $H^0$.

\section{Higgs production analysis\label{higgprodanalysis}}
We now consider the different processes at an $e^+ e^-$ collider which can be
 used for the measurements of the trilinear Higgs couplings
 $\lambda_{hhh}$ and $\lambda_{Hhh}$ in the MSSM. These 
processes involve production of multiple Higgs bosons, to which we now turn.

\begin{figure}[t]
\centering
\includegraphics[trim=50 650 0 0, clip]{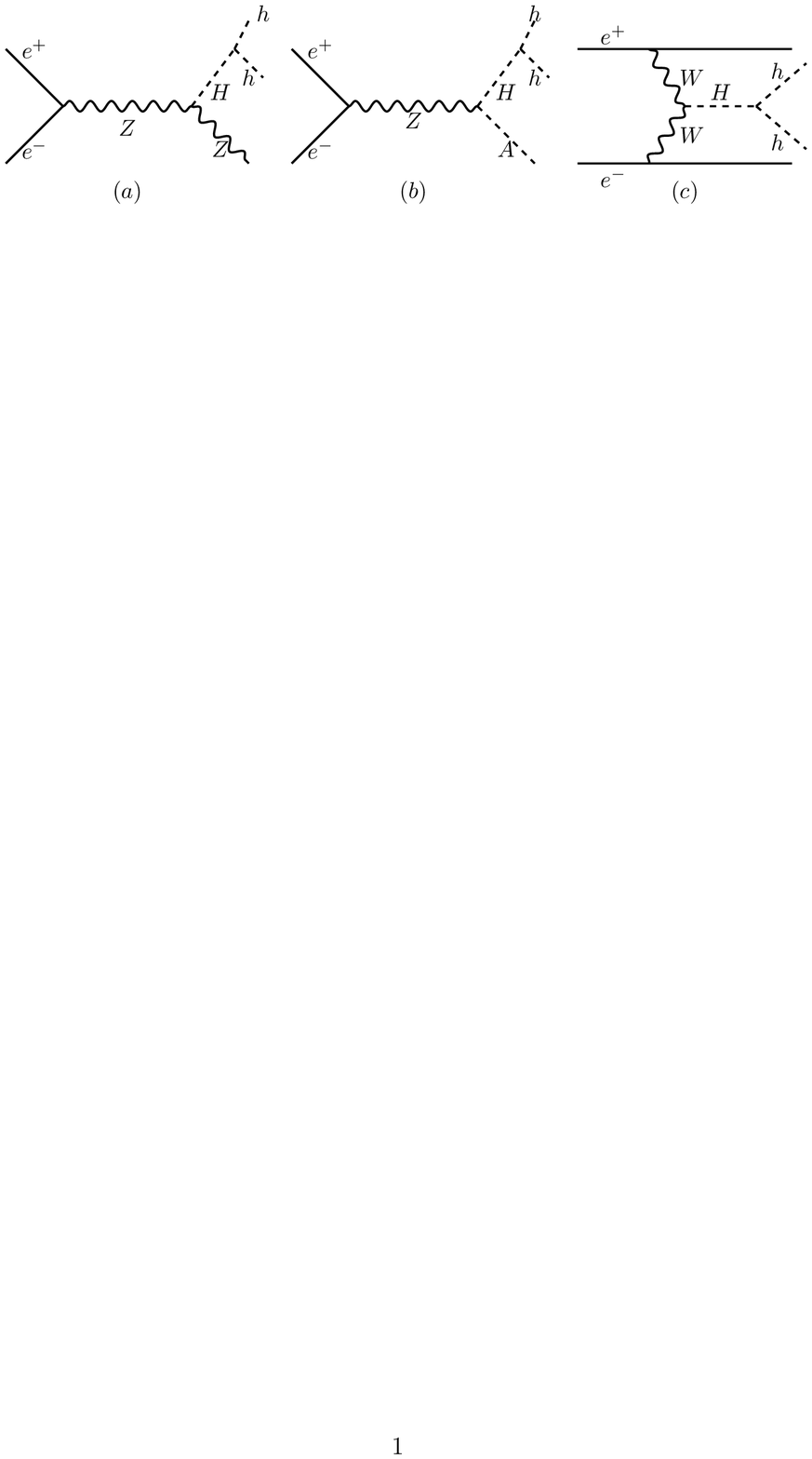}
\caption{Feynman diagrams for resonant production
 of  $hh$, through $e^+ e^-$ $\rightarrow$
$HZ$, $HA$, $\nu_e \bar\nu_e H $, where $H \rightarrow hh$ in the
final state.}
\vspace{1.0cm}\label{feyn1}
\end{figure}

\begin{figure}[htb]
 \centering
\includegraphics[trim=50 600 0 0, clip]{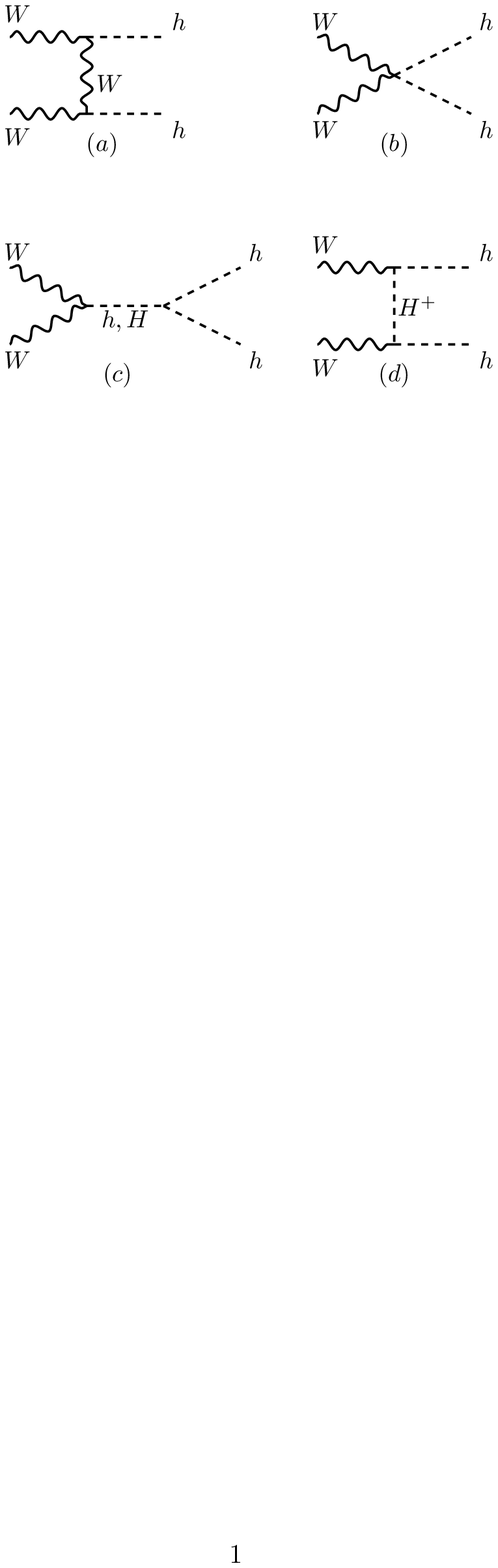}
\vspace{.5cm} \caption{Feynman diagrams for $hh$
production through non-resonant $WW$ fusion.} \vspace{1.0cm} \label{feyn2}
\end{figure}

Multiple light Higgs bosons ($h^0$) can be  produced through heavy
CP-even Higgs boson decays. For CP-even
 heavy Higgs boson production we consider Higgs-strahlung
$e^+ e^- \rightarrow ZH $, associated production with CP-odd Higgs boson $e^+
e^- \rightarrow AH $, and $WW$ fusion mechanism $e^+ e^-
\rightarrow \nu_e \bar\nu_e H $. Feynman diagrams for these processes are shown in Figs.
\ref{feyn1} and \ref{feyn2}. Heavy Higgs boson $H^0$ subsequently
 decays to a pair of light Higgs bosons. The branching ratio of $H
\rightarrow hh$ depends on  trilinear Higgs coupling $\lambda_{Hhh}$,
\be  \Gamma(H \rightarrow hh) = \frac{G_F \lambda_{Hhh}^2}{16 \sqrt{2} \pi} 
\frac{m_Z^4}{m_{H^0}} (1-4 \frac{m_{h^0}^2}{m_{H^0}^2})^{1/2}. \ee
Notice that this decay is kinematically forbidden in the non decoupling regime.
The cross-sections for the Higgs-strahlung and
 associated production with CP-odd Higgs boson  can be written
 as \cite{Pocsik:1981bg,Gunion:1988tf,OPtricooup,Osland:1999ae,Osland:1999qw}

\begin{eqnarray}
\sigma (e^+e^- \rightarrow ZH) & = & \frac{G_F^2m_Z^4}{96\pi s}
(v_e^2 + a_e^2)\cos^2(\beta - \alpha) 
\frac{\lambda_Z^{1/2} \left [\lambda_Z + 12m_Z^2/s
\right ]}{(1-m_Z^2/s)^2}, \label{Eq:sigZH}\\
\sigma (e^+e^- \rightarrow AH) & = & \frac{G_F^2m_Z^4}{96\pi s}
(v_e^2 + a_e^2)\sin^2(\beta - \alpha) 
\frac{\lambda_A^{3/2}}{(1-m_Z^2/s)^2}, 
\label{Eq:sigAH}
\end{eqnarray}
where $\lambda_i$ is the phase-space function, which corresponds
 to $\lambda(m_i^2, m_{H^0}^2; s)$, 
and is given by
\begin{equation}
\lambda(m_a^2, m_b^2; m_c^2) = \left(1 - \frac{m_a^2}{m_c^2} - 
\frac{m_b ^2}{m_c^2}\right)^2 - \frac{4 m_a^2m_b^2}{m_c^4},
\label{27}
\end{equation}
and $v_e=4 \sin^2 \theta_W -1$, $a_e=-1$ are the Z boson-electron couplings.

%\noindent
On the other hand resonant $WW$ fusion cross-section for the light Higgs pair
 production can be written as \cite{OPtricooup,Osland:1999ae,Osland:1999qw}
\begin{equation}
\sigma(e^+e^- \rightarrow H\bar\nu_e\nu_e) 
= \frac{G_F^3 m_W^4}{64 \sqrt 2\pi^3 }
\left[\int_{\mu_H}^1 dx\int_{x}^1 \frac{dy}
{\left[1 + (y-x)/\mu_W\right]^2}\ {\cal F}(x,y)\right]
\cos^2(\beta - \alpha), 
\label{Eq:fusion-exact}
\end{equation}
where 
\begin{eqnarray}
{\cal F}(x,y)
& = & 16[F(x,y)+G(x,y)], \\
F(x,y) & = & \left[ \frac{2x}{y^3} - \frac{1 + 2x}{y^2}
+\frac{2 + x}{2y} - \frac{1}{2}\right]
\left[\frac{z}{1 + z} - \log(1 + z)\right]
+ \frac{x}{y^3}\frac{z^2(1 - y)}{(1 + z)}, \\ 
G(x,y) & = & \left[ -\frac{x}{y^2} 
+\frac{2 + x}{2y} - \frac{1}{2}\right]
\left[\frac{z}{1 + z} - \log(1 + z)\right],
\end{eqnarray}
with 
\begin{equation}
\mu_{H}=\frac{m_{H^0}^2}{s},\quad \quad z = \frac{y(x - \mu_H)}{\mu_W x}.
\end{equation}
The multiple Higgs production through non resonant $WW$ $\rightarrow$ $hh$ proceeds
 via the diagrams shown in the Fig. \ref{feyn2}.
The non-resonant fusion $WW \rightarrow hh$ cross-section in
 the effective $WW$ approximation can be written as 
\begin{equation}
\label{Eq:sigWW-nonres}
\sigma(e^+e^-\to\nu_e\bar\nu_e hh)
=\int_\tau^1\dd x\, \frac{\dd L}{\dd x}\, \hat\sigma_{WW}(x),
\end{equation}
where 
\begin{eqnarray}
\tau & = & \frac{4m_{h^0}^2}{s}, \\
\frac{\dd L(x)}{\dd x}
& = & \frac{G_{\rm F}^2m_W^4}{2}\,\left(\frac{v^2+a^2}{4\pi^2}\right)^2
\frac{1}{x}\biggl\{(1+x)\log\frac{1}{x}-2(1-x)\biggr\},
\end{eqnarray}
and $\hat\sigma_{WW}(x)$ can be written as \cite{OPtricooup,Osland:1999ae,Osland:1999qw}

\begin{eqnarray}
\label{Eq:sighat}
\hat\sigma_{WW}(x)
&=&\frac{G^2_{\rm F}\hat s}{64\pi}\beta_h
\biggl\{4\biggl[
 \frac{\hat\mu_Z\sin(\beta-\alpha)}{1-\hat\mu_h}\,\lambda_{hhh}
+\frac{\hat\mu_Z\cos(\beta-\alpha)}{1-\hat\mu_H}\,\lambda_{Hhh}
+1\biggr]^2\,g_0 \nonumber \\
&& \phantom{\frac{2}{\beta_h}}
+\frac{2}{\beta_h}\biggl[
 \frac{\hat\mu_Z\sin(\beta-\alpha)}{1-\hat\mu_h}\,\lambda_{hhh}
+\frac{\hat\mu_Z\cos(\beta-\alpha)}{1-\hat\mu_H}\,\lambda_{Hhh}
+1\biggr]\nonumber \\
&& \phantom{\frac{2}{\beta_h}AAAA}
\times[\sin^2(\beta-\alpha)\,g_1 +\cos^2(\beta-\alpha)\,g_2] \nonumber \\
&& \phantom{\frac{2}{\beta_h}}
+\frac{1}{\beta_h^2}
\{\sin^4(\beta-\alpha)\,g_3+\cos^4(\beta-\alpha)\,g_4
+\sin^2[2(\beta-\alpha)]\,g_5\}\biggr\},
\end{eqnarray}
where 
\begin{equation}
\hat\mu_i=\frac{m_i^2}{\hat s} \,\,(i=W,Z,h^0,H^0), \quad
\beta_h=\left(1-4\hat\mu_h\right)^{1/2}, \quad \hat s= xs,
\end{equation}
and the exact forms of $g_i$ (i=0,..5) functions can be found 
in \cite{OPtricooup,Osland:1999ae,Osland:1999qw}. We note that there can be sizable deviations
 of the effective $WW$ approximation from the exact result. However,
 we shall use this approximation as an estimate in this paper.

\begin{figure}[htb]
 \centering
\includegraphics[trim=50 595 0 0, clip]{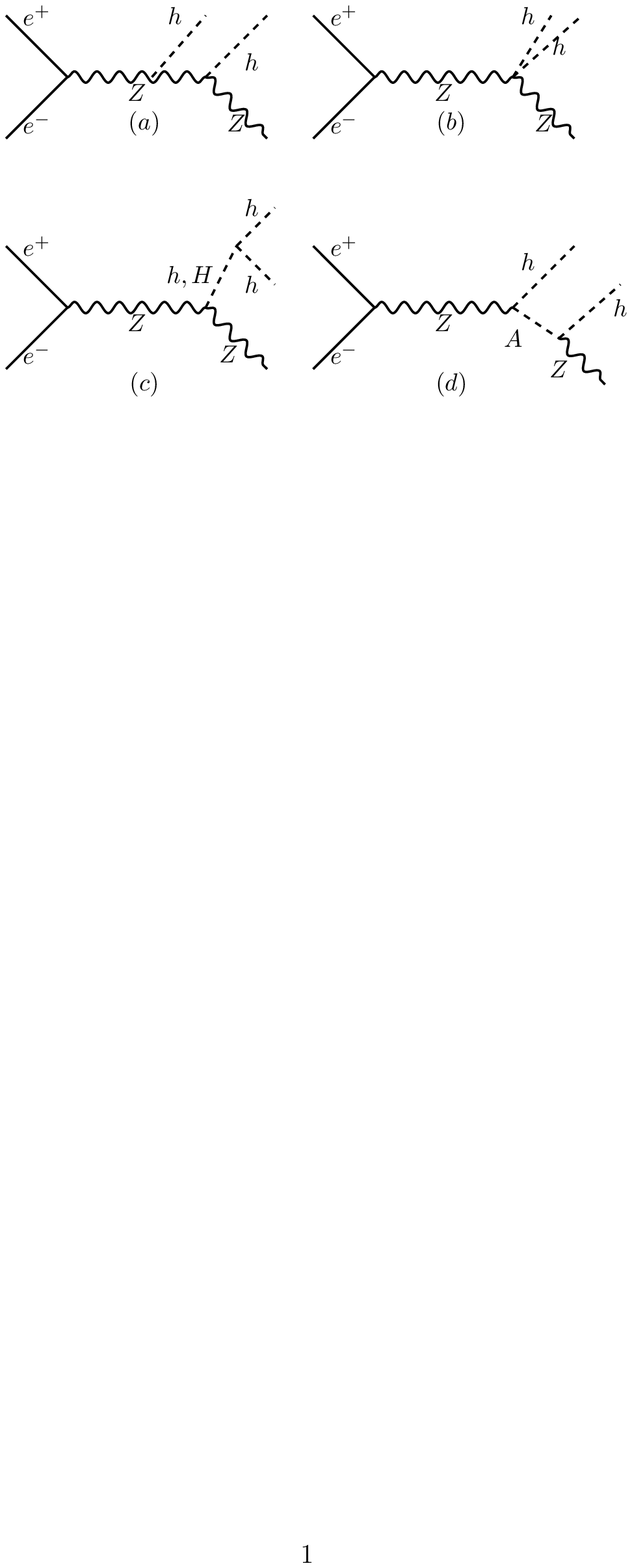}
\vspace{.5cm} \caption{Feynman diagrams for the non-resonant $hh$
production in association with $Z$.} \vspace{1.0cm} \label{feyn3}
\end{figure}

\begin{figure}[htb]
 \centering
\includegraphics[trim=50 650 0 0, clip]{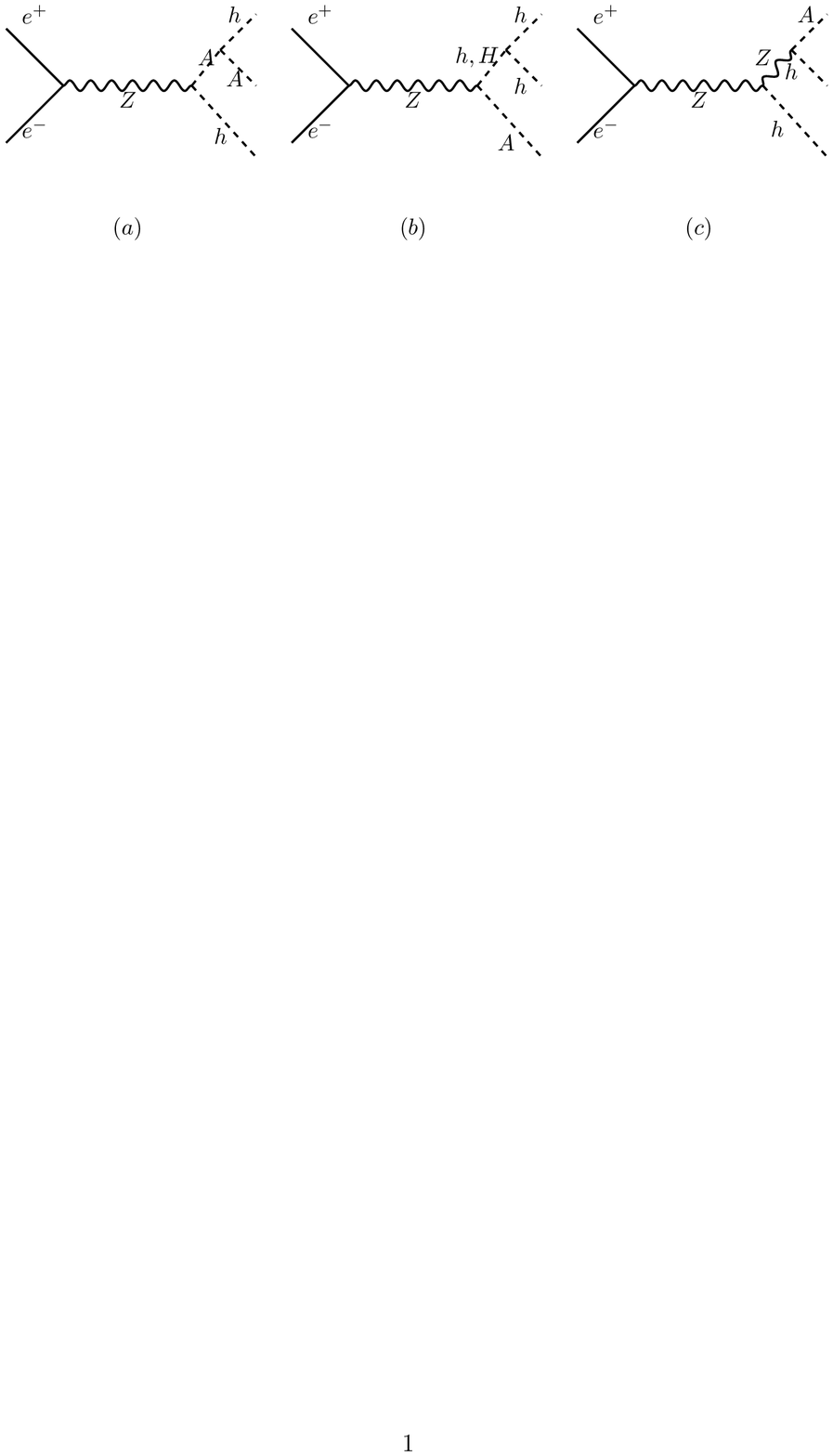}
\vspace{.5cm} \caption{Feynman diagrams for the non-resonant $hh$
production in association with $A$.} \vspace{1.0cm} \label{feyn4}
\end{figure} \noindent
The off-shell $Z$ boson
decay 
\begin{figure}[t]
\centering \subfigure[]{
\includegraphics[scale=0.41]{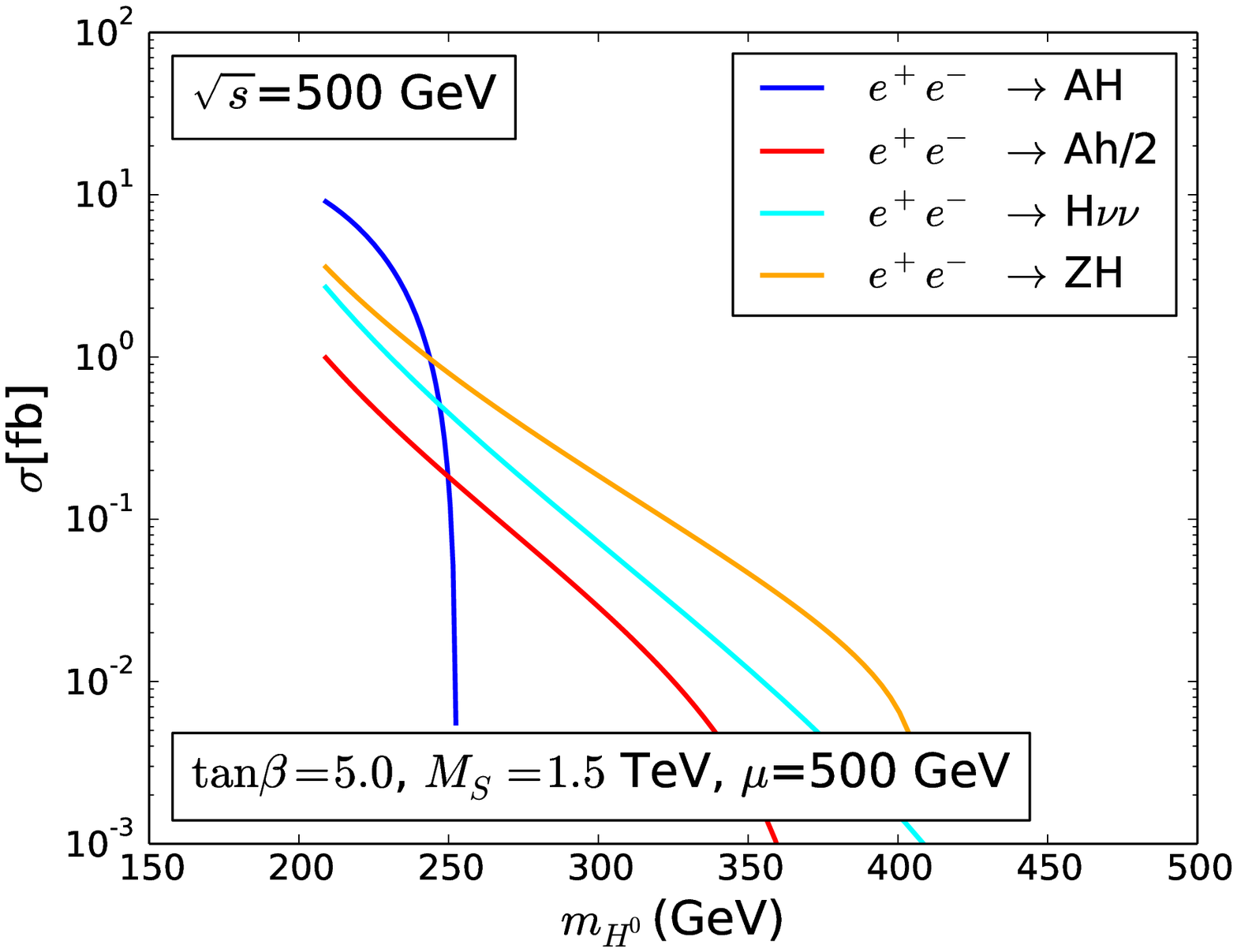}}
\hspace{-.8cm}\centering \subfigure[]{
\includegraphics[scale=0.41]{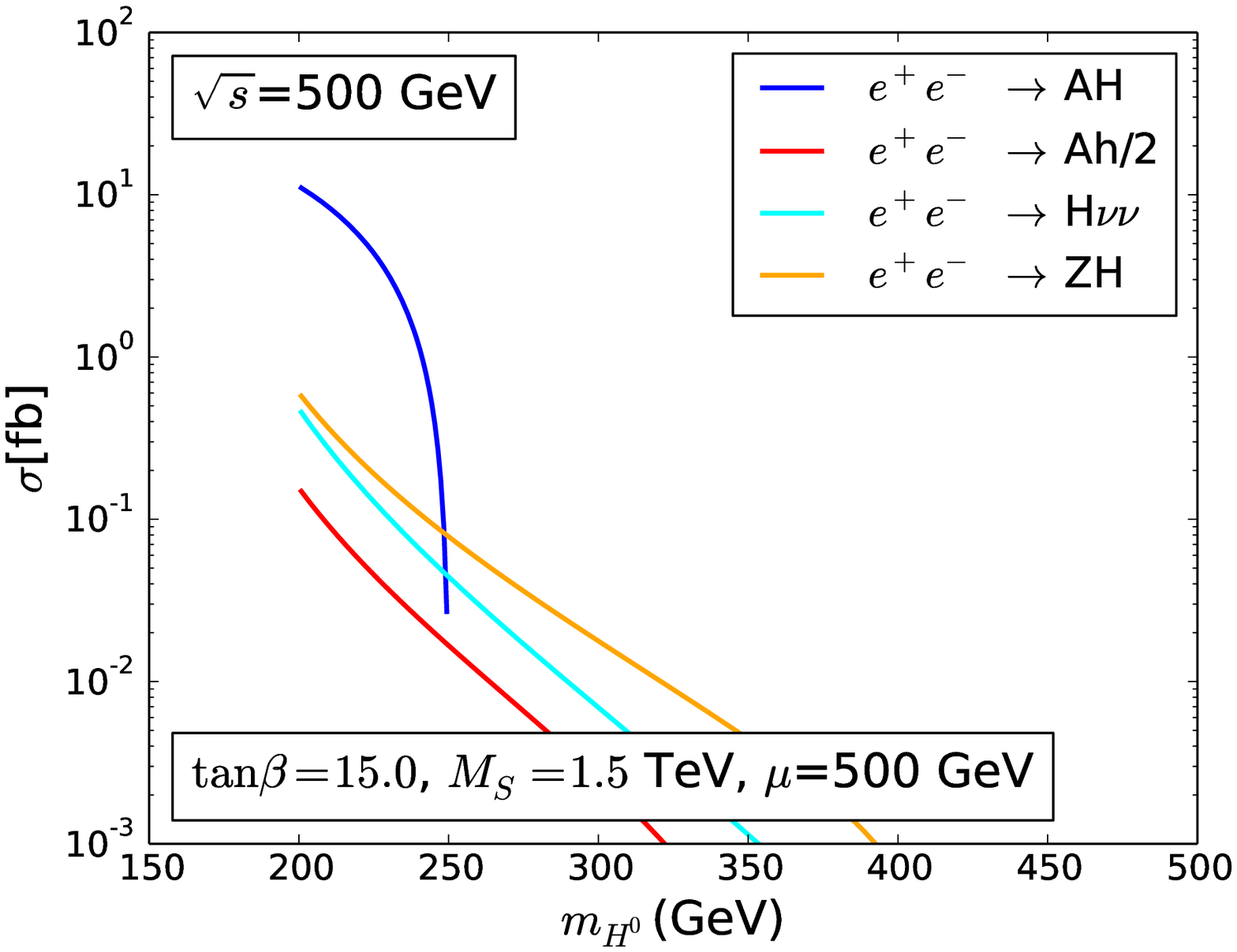}}
\centering \subfigure[]{
\includegraphics[scale=0.41]{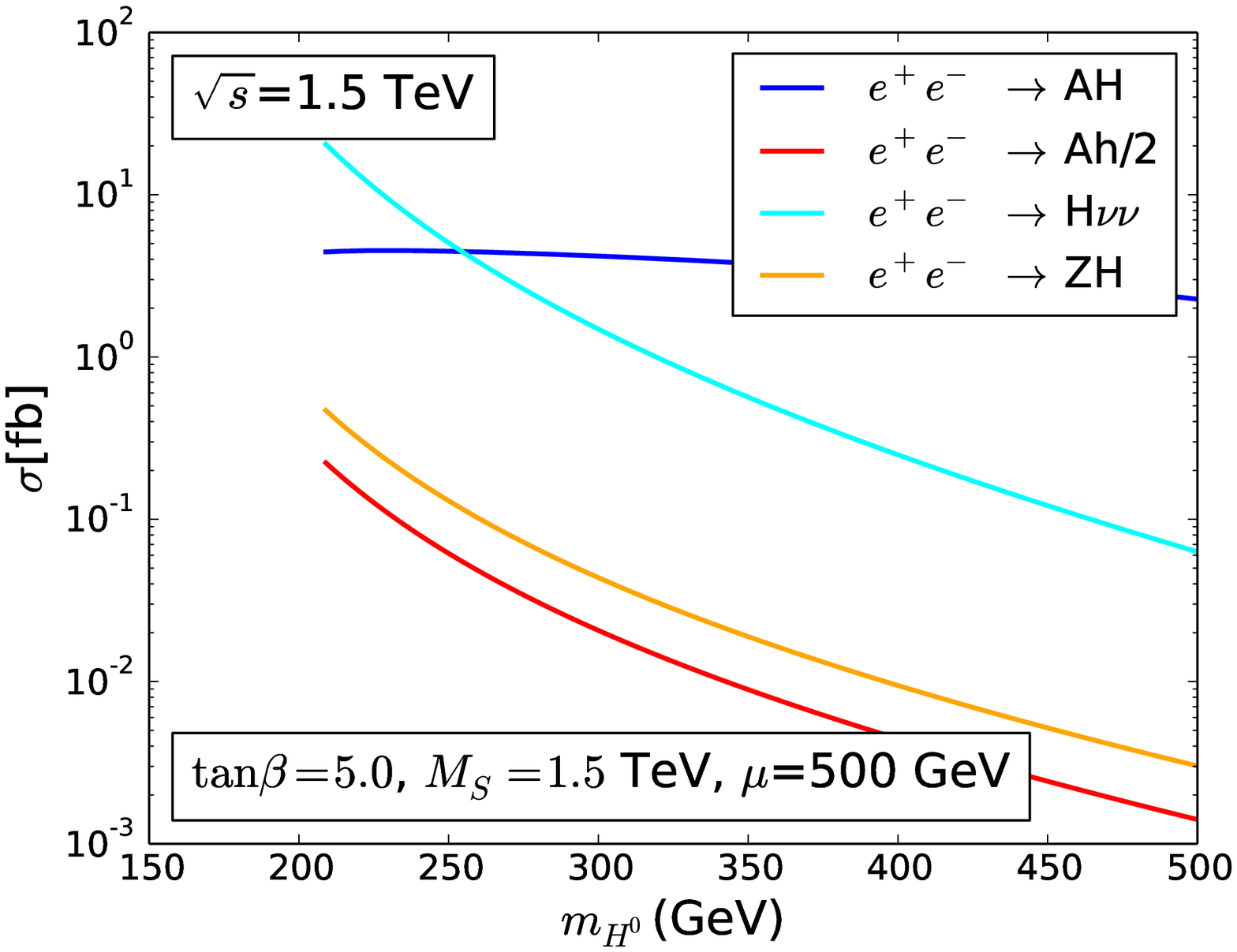}}
\centering \subfigure[]{
\hspace{-.8cm}\includegraphics[scale=0.41]{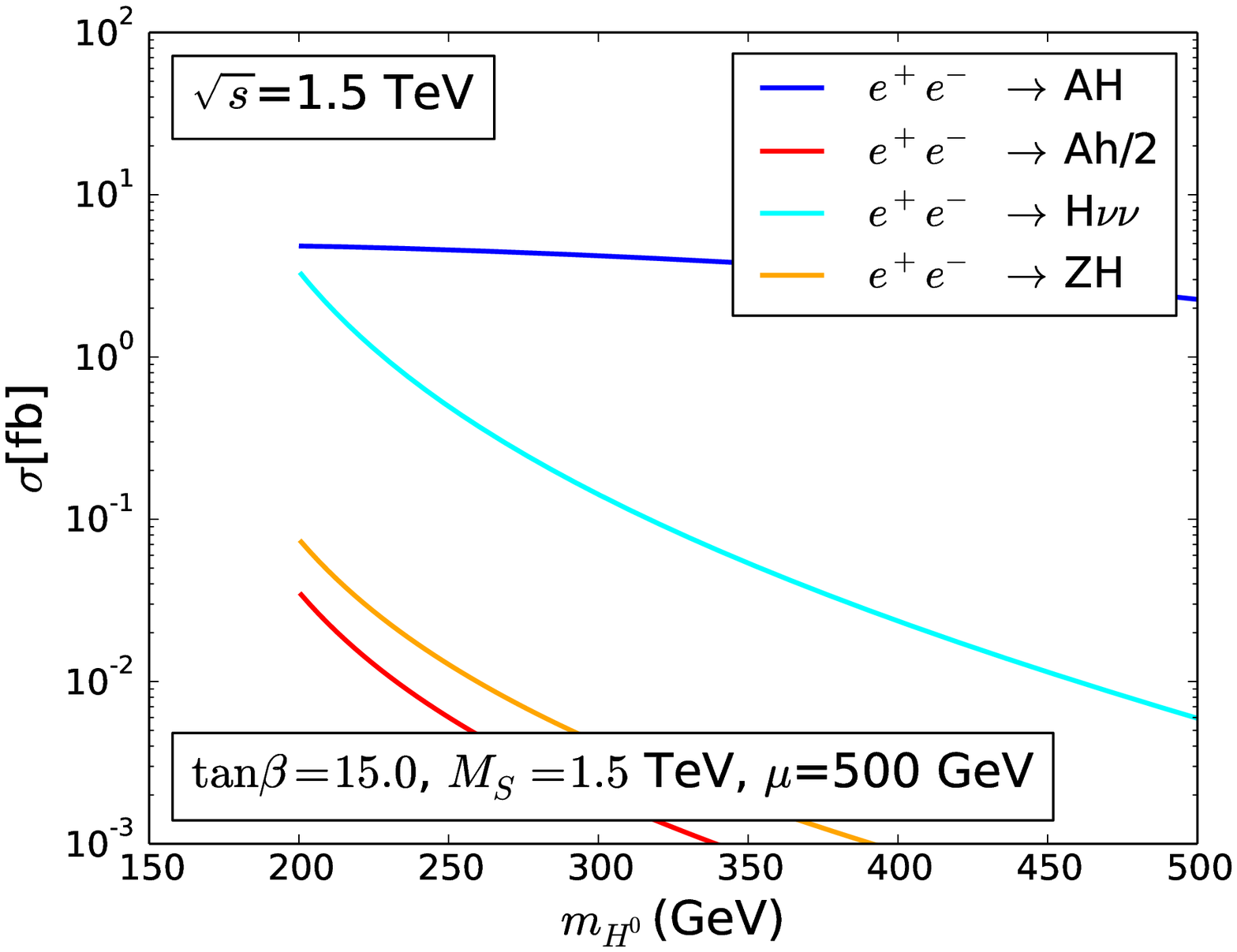}}
\caption{Cross-section for $e^+ e^- \rightarrow AH, ZH, Ah/2, H\nu\nu$ as
 the functions of $m_{H^0}$ for $\sqrt{s}$ = 500 GeV
and  $\sqrt{s}$ = 1.5 TeV for values of other
 parameters as shown in the inset. Soft trilinear
 coupling $A_t$ is adjusted to
get light Higgs boson mass in 122-128 GeV range. 
}\label{crossplots}
\end{figure}
\be   e^+ e^- \rightarrow Z^* \rightarrow Zhh  \label{hhwithzprod}, \ee
is another mechanism of $hh$ production.
Feynman diagrams for this process are shown in Fig. \ref{feyn3}, and
 the cross-section is given by \cite{OPtricooup,Osland:1999ae,Osland:1999qw} 
\begin{equation} 
\frac{d\sigma(e^+e^- \rightarrow Zhh)}{dx_1 dx_2}
= \frac{G_F^3 m_Z^6}{384\sqrt 2 \pi^3s}(v_e^2 + a_e^2) 
\frac{\mathcal A}{(1-\mu_Z)^2},
\label{Eq:sigZhh}
\end{equation}
where  $x_{1,2} = {{2E_{1,2}}/{\sqrt s}}$
are the scaled energies of the Higgs particles, $x_3 = 2 - x_1 -x_2$
for the scaled energy of the $Z$ boson, and $y_k = 1 - x_k$. 
Also, $\mu_i = m_i^2/s$  denotes the scaled squared masses of various 
particles:
\begin{equation}
\mu_h = m_{h^0}^2/s, \qquad  \mu_H = m_{H^0}^2/s, \qquad \mu_W = m_W^2/s,
\label{Eq:mu}
\end{equation}
and
\begin{equation}
{\mathcal A}
 = \mu_Z\, \left\{ \half|a|^2\, f_a
+|b(y_1)|^2\, f_b +2\, \Re[a b^*(y_1)]\, g_{ab}
+\Re[b(y_1)b^*(y_2)]\,g_{bb}\right\} +\{x_1\leftrightarrow x_2\}.
\label{Eq:calA}
\end{equation}
Here
\begin{equation}
a = \frac{1}{2}
\biggl[\frac{\sin(\beta-\alpha)\lambda_{hhh}}{y_3+\mu_Z-\tilde\mu_h}
+\frac{\cos(\beta-\alpha)\lambda_{Hhh}}{y_3+\mu_Z-\tilde\mu_H}\biggr]
+\biggl[\frac{\sin^2(\beta-\alpha)}{y_1+\mu_h-\tilde\mu_Z}
+\frac{\sin^2(\beta-\alpha)}{y_2+\mu_h-\tilde\mu_Z}\biggr]
+\frac{1}{2\mu_Z},
\label{Eq:lca}
\end{equation}
\begin{equation}
b(y) = \frac{1}{2\mu_Z}\left(
 \frac{\sin^2(\beta-\alpha)}{y+\mu_h-\tilde\mu_Z}
+\frac{\cos^2(\beta-\alpha)}{y+\mu_h-\tilde\mu_A}
\right).
\label{Eq:lcb}
\end{equation}
and  $\tilde\mu_Z=(m_Z^2+im_Z\Gamma_Z)/s$, which takes care
 of the widths. The Higgs self-couplings
 $\lambda_{Hhh}$ and $\lambda_{hhh}$ occur only
in the function $a$, Eq.~(\ref{Eq:lca}). 
The coefficients $f$ and $g$
do not involve any Higgs couplings and can be written as
\begin{eqnarray}
f_a &=&  x_3^2+8\mu_Z, \nonumber \\
f_b &=&  (x_1^2-4\mu_h)[(y_1-\mu_Z)^2-4\mu_Z\mu_h], \nonumber \\
g_{ab} &=&  \mu_Z[2(\mu_Z-4\mu_h)+x_1^2+x_2(x_2+x_3)]
-y_1(2y_2-x_1x_3),\nonumber \\
g_{bb} &=& \mu_Z^2(4\mu_h +6 -x_1x_2) +2\mu_Z(\mu_Z^2 +y_3 -4\mu_h) 
\nonumber \\
& & +(y_3-x_1x_2-x_3\mu_Z-4\mu_h\mu_Z)(2y_3-x_1x_2-4\mu_h+4\mu_Z).
\end{eqnarray}
\noindent
  We note that the
 Feynman diagram Fig. \ref{feyn3}(c) involves the trilinear Higgs couplings
 $\lambda_{Hhh}$ and $\lambda_{hhh}$, whereas the other diagrams
 in Fig. \ref{feyn3} do not involve any trilinear Higgs couplings.
The background to the multiple Higgs production process comes from pseudoscalar $A$ production
 with $h$, where $A$ subsequently decays to
 $hZ$ (see Fig. \ref{feyn3}(d) for the corresponding Feynman diagram)
\be  e^+ e^- \rightarrow A h, \qquad  \qquad A \rightarrow h Z.   \ee
 The production
 cross-section for $e^+ e^-$ $\rightarrow$ $Ah$ can be written as 

\begin{equation}
\sigma(e^+e^- \rightarrow Ah)  =  \frac{G_F^2 m_Z^4}{96\pi s}
(v_e^2 + a_e^2) \cos^2(\beta - \alpha)
\frac{\lambda^{3/2}(m_{h^0}^2, m_{A^0}^2; s)}{(1 - m_Z^2/ s)^2}, 
\label{28}
\end{equation}
and decay width for $A \rightarrow h Z$ is given by 
 \begin{equation}
\Gamma(A \rightarrow hZ) =  \frac{G_F}{8\pi\sqrt 2}
\cos^2(\beta - \alpha) \frac{m_Z^4}{m_{A^0}}
\lambda^{1/2}(m_Z^2, m_{h^0}^2; m_{A^0}^2) \lambda(m_{A^0}^2, m_{h^0}^2; m_Z^2).
\label{26}
\end{equation}
We note that the Feynman diagrams shown in Fig. \ref{feyn4} will lead to
 $hhh$ final state through $A \rightarrow h Z$ decay, whereas we are
 interested in final states having $hh$ final state.

In Fig. \ref{crossplots} we show the  cross-section for $e^+ e^-$ $\rightarrow$
$HZ$, $HA$,  $H \nu \bar\nu$, (Fig.\ref{feyn1}), $Ah $ (Fig. \ref{feyn3}(d))
 as a function of $m_{H^0}$ for different values of $\sqrt{s}$ and
 $\tan\beta$. The heavy Higgs production with CP-odd Higgs 
is the dominant production channel for $m_{H^0}$ $\le$ 250 GeV. We
 can see that  ($\sigma(e^+ e^- \rightarrow Ah)$)/2 is of order
 of $\sigma(e^+ e^- \rightarrow ZH)$  and $\sigma(e^+ e^- \rightarrow H \nu \bar\nu)$.

\begin{table}
 $$
 \begin{array}{|c|c|c|c|c|}
\hline{ {\mu}} &  {M_1} & {M_2}&  {M_{\chi^0}} & {M_{\chi^{\pm}}}\\
{\small{ \mbox{(in GeV)}}} & {\small{ \mbox{(in GeV)}}} &
 {\small{ \mbox{(in GeV)}}}& {\small{ \mbox{(in GeV)}}} & {\small{ \mbox{(in GeV)}}} \\
 \hline
 -230.0   &      120.0   &
 240.0 &108.0, 181.7, 235.6, 305.9  & 173.5, 304.1\\ \hline
-500.0   &      150.0   &
 300.0 &146.9, 283.4, 503.2, 522.9  & 282.9, 521.6\\ \hline
  \end{array}
 $$
\caption{The neutralino and chargino mass spectrum for the benchmark
 values of the $\mu$ parameter and soft supersymmetric
 breaking gaugino masses $M_1$ and $M_2$.}
 \label{table1} \end{table}

\begin{figure}[t]
\centering \subfigure[]{
\includegraphics[scale=0.41]{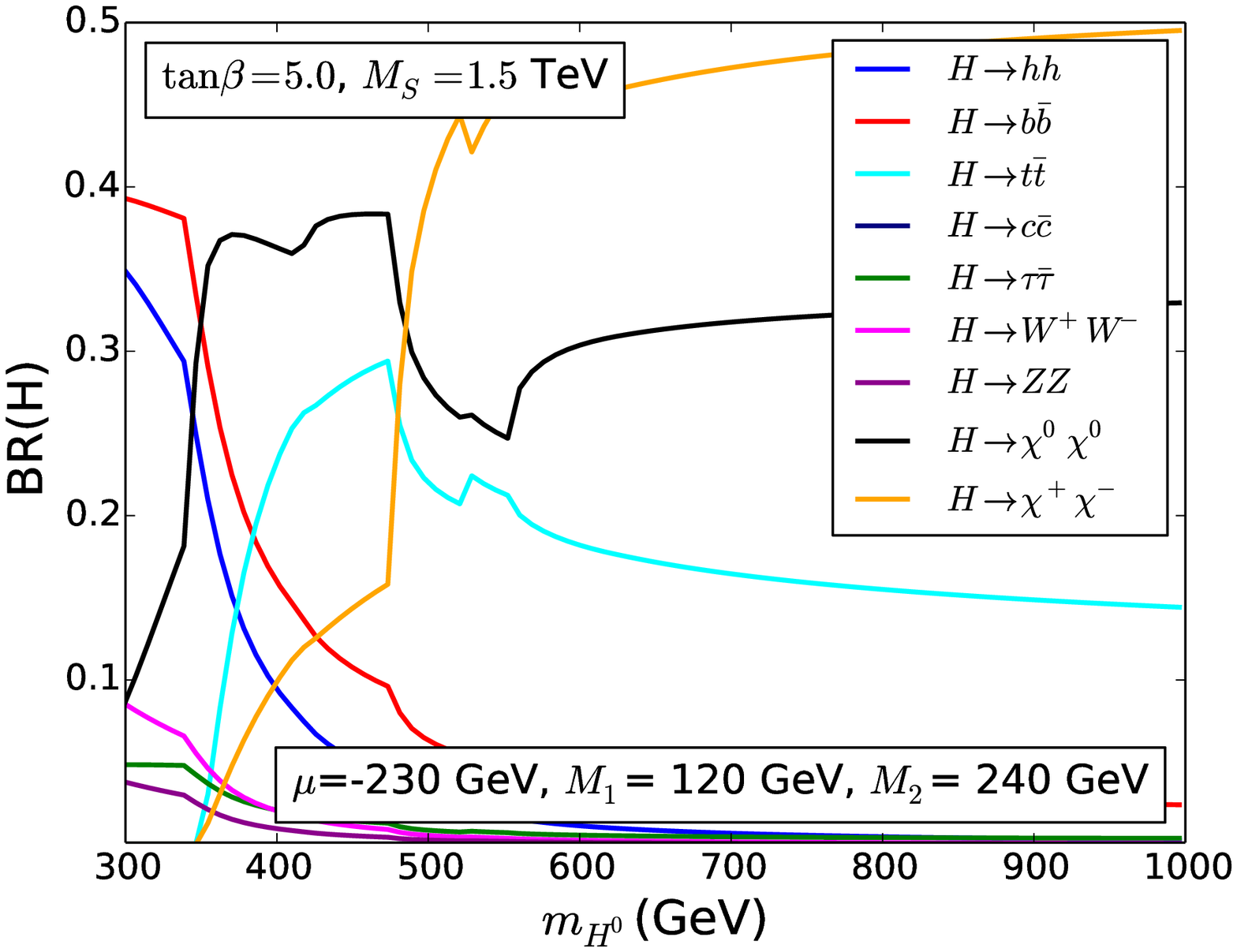}}
\hspace{-.8cm}\centering \subfigure[  ]{
\includegraphics[scale=0.41]{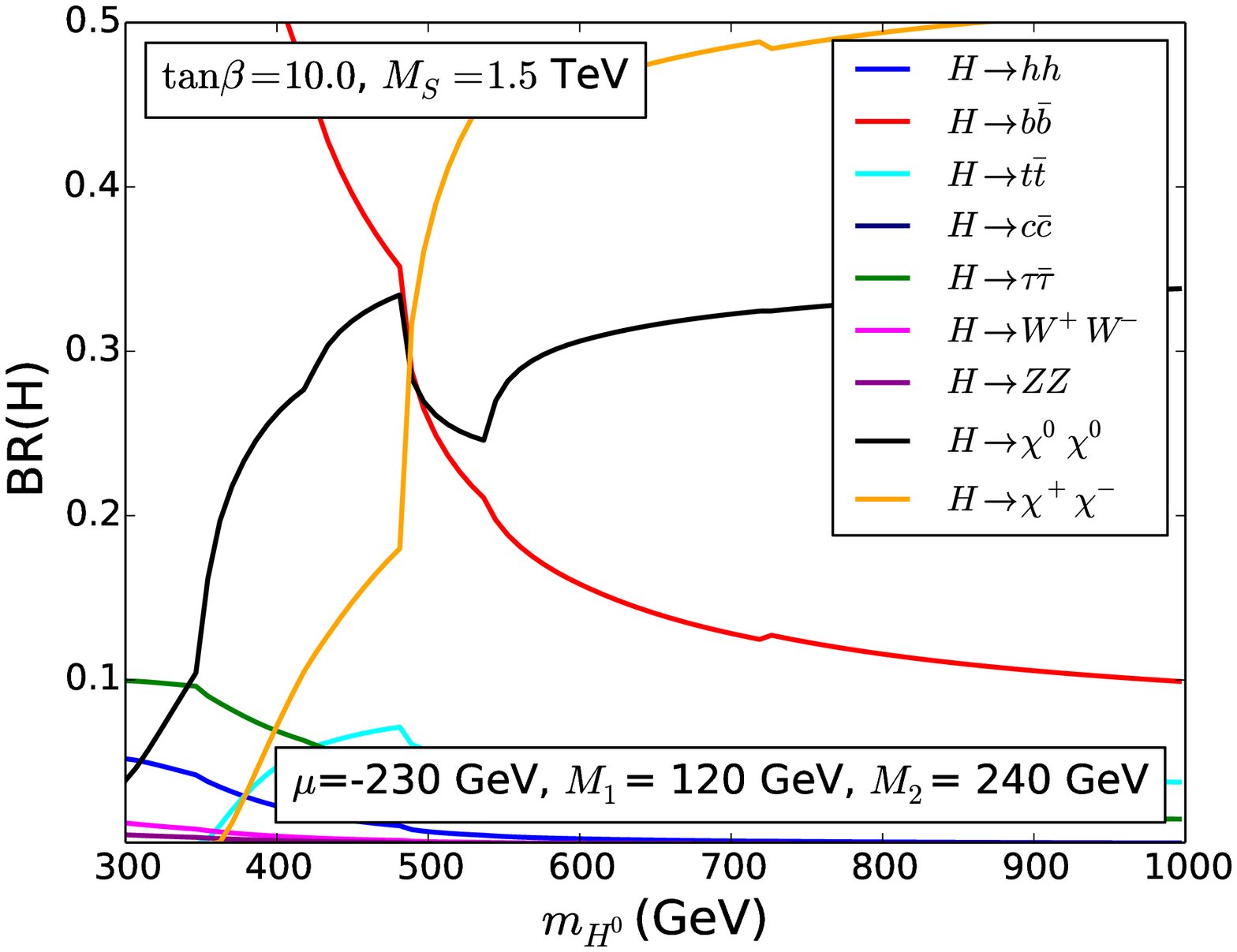}}
\centering \subfigure[ ]{
\includegraphics[scale=0.41]{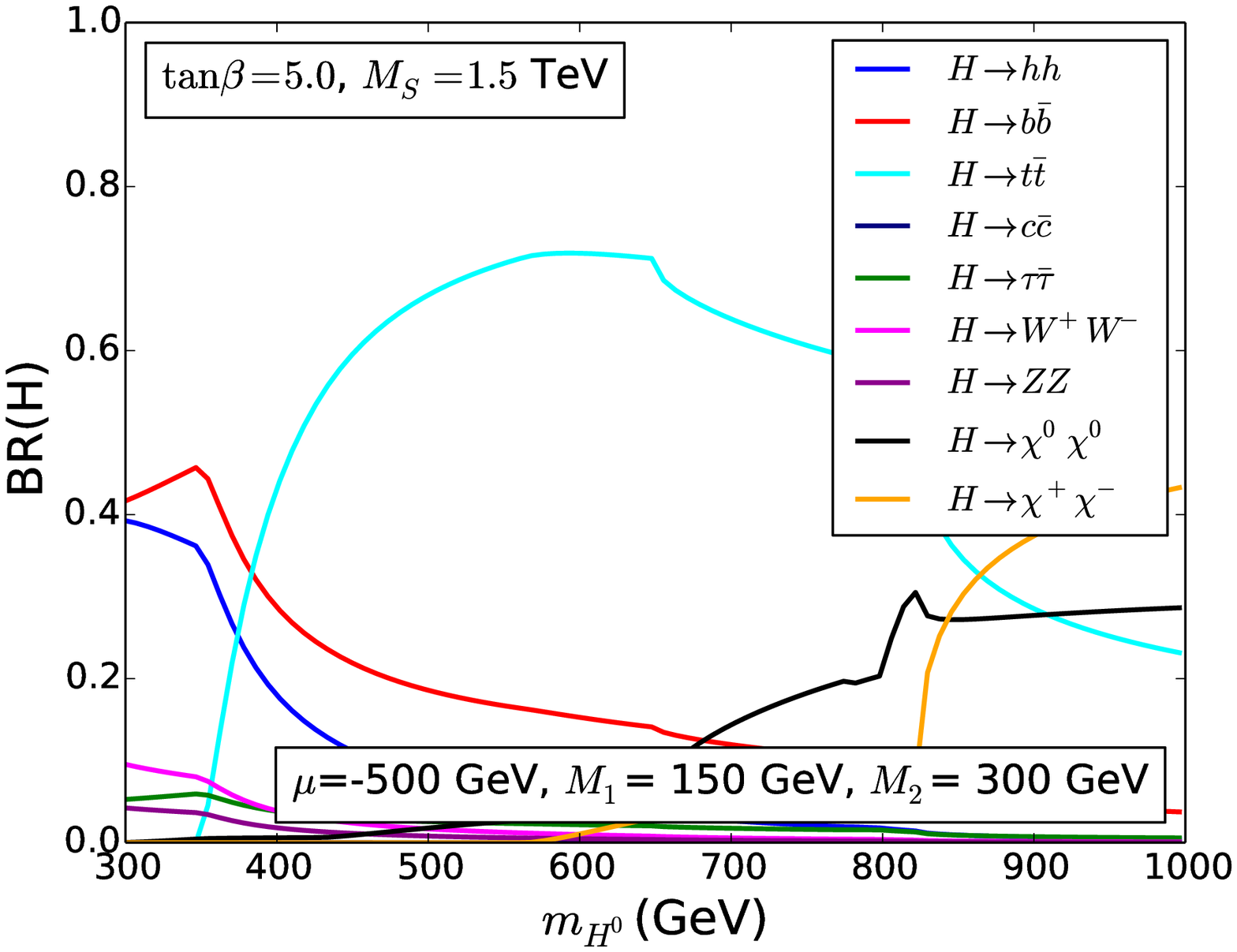}}
\hspace{-.8cm}\centering \subfigure[]{
\includegraphics[scale=0.41]{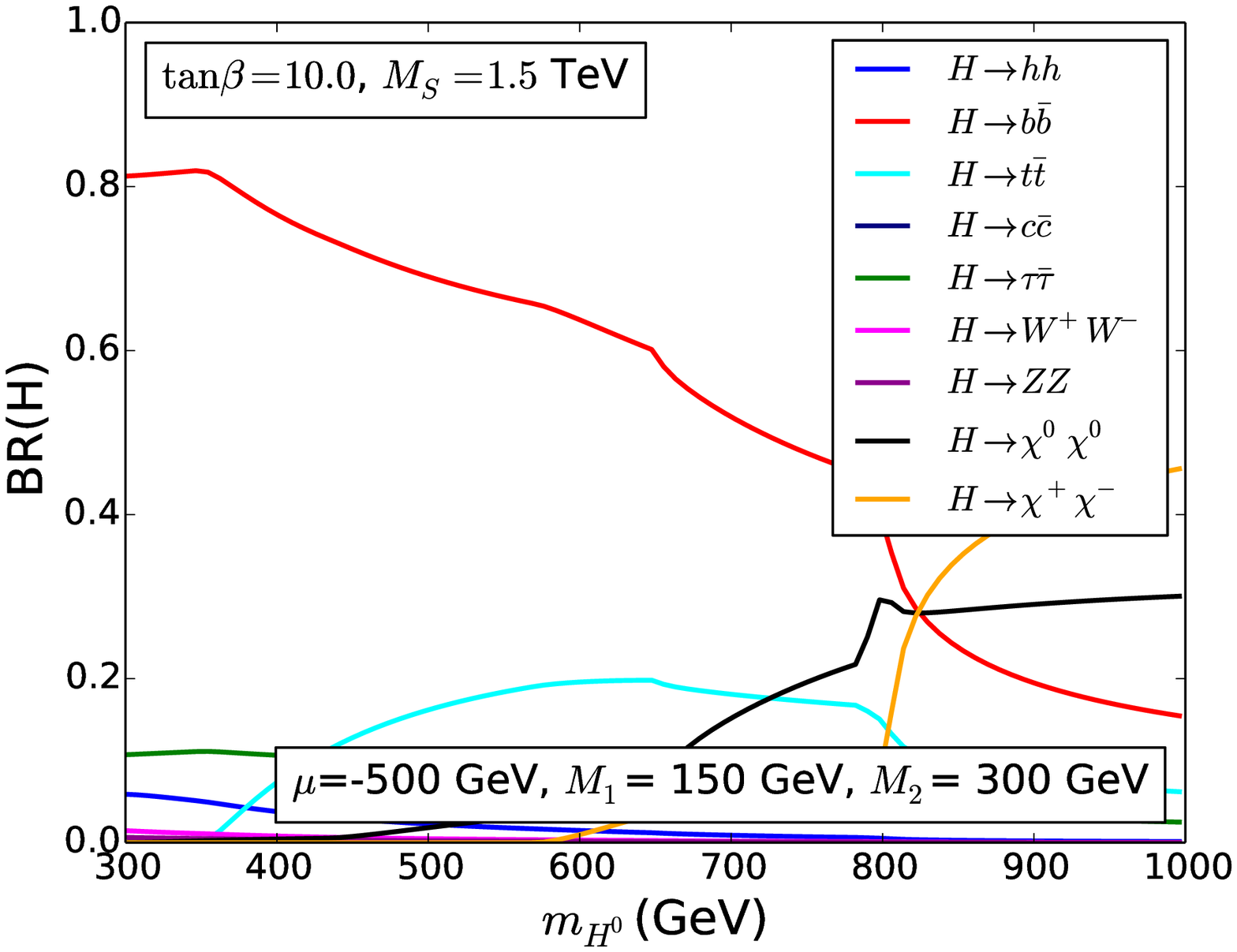}}
\caption{The branching fractions for different decay
 channels of  heavy Higgs (H) boson as a function of $m_{H^0}$, 
 for the fixed values of parameters shown in the inset.}\label{hdecay}
\end{figure}

\begin{figure}[t]
\centering \subfigure[]{
\includegraphics[scale=0.41]{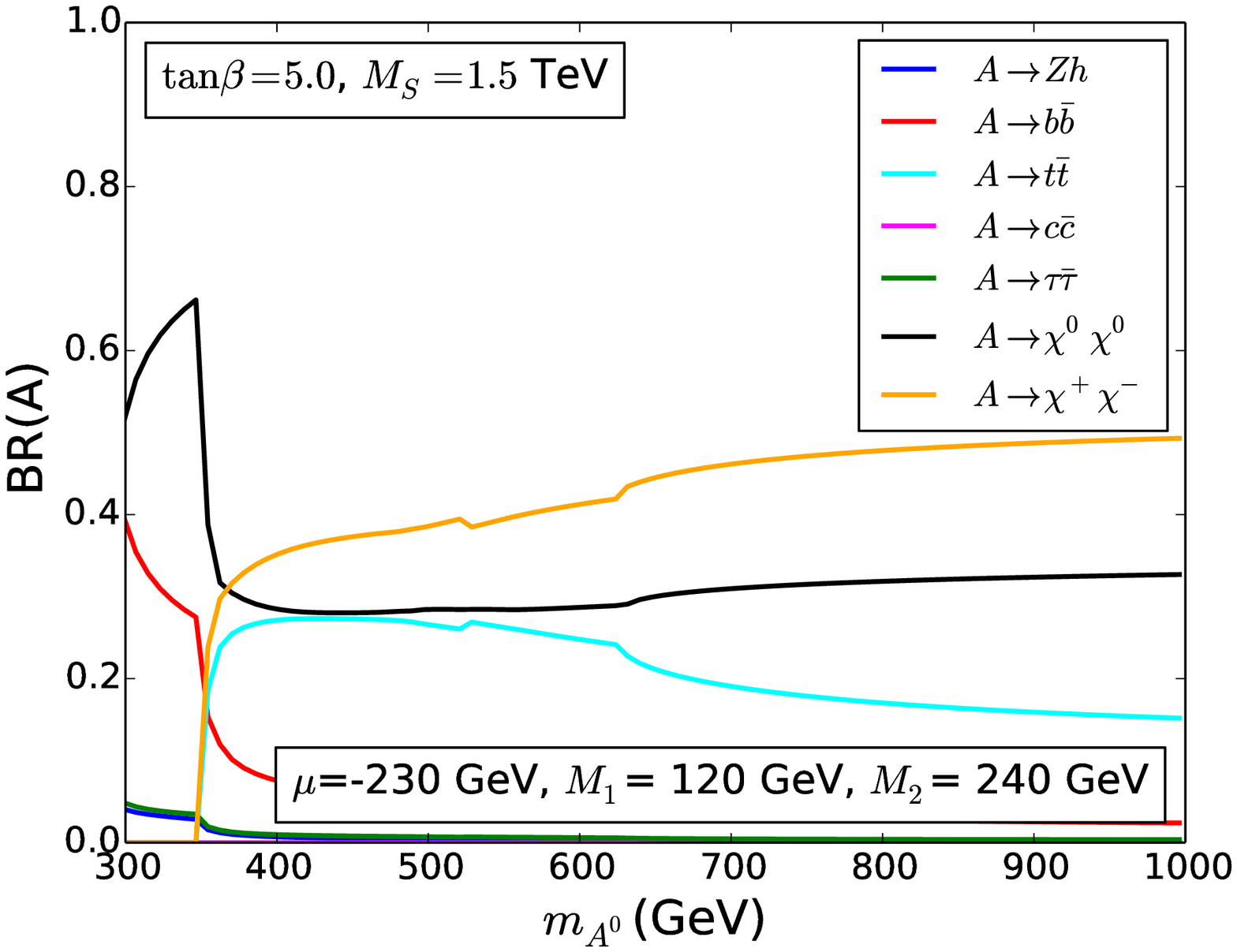}}
\hspace{-.8cm}\centering \subfigure[]{
\includegraphics[scale=0.41]{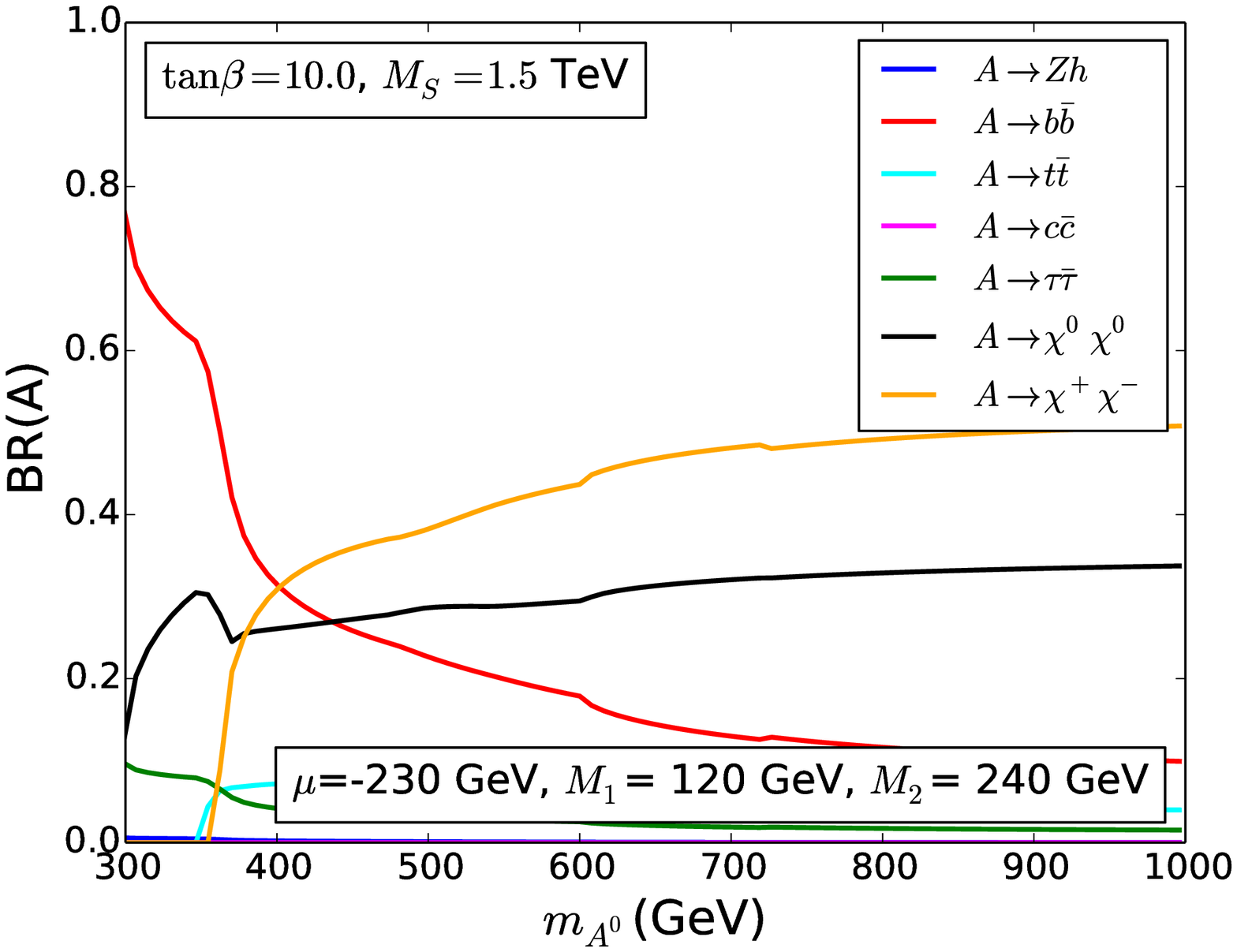}}
\centering \subfigure[]{
\includegraphics[scale=0.41]{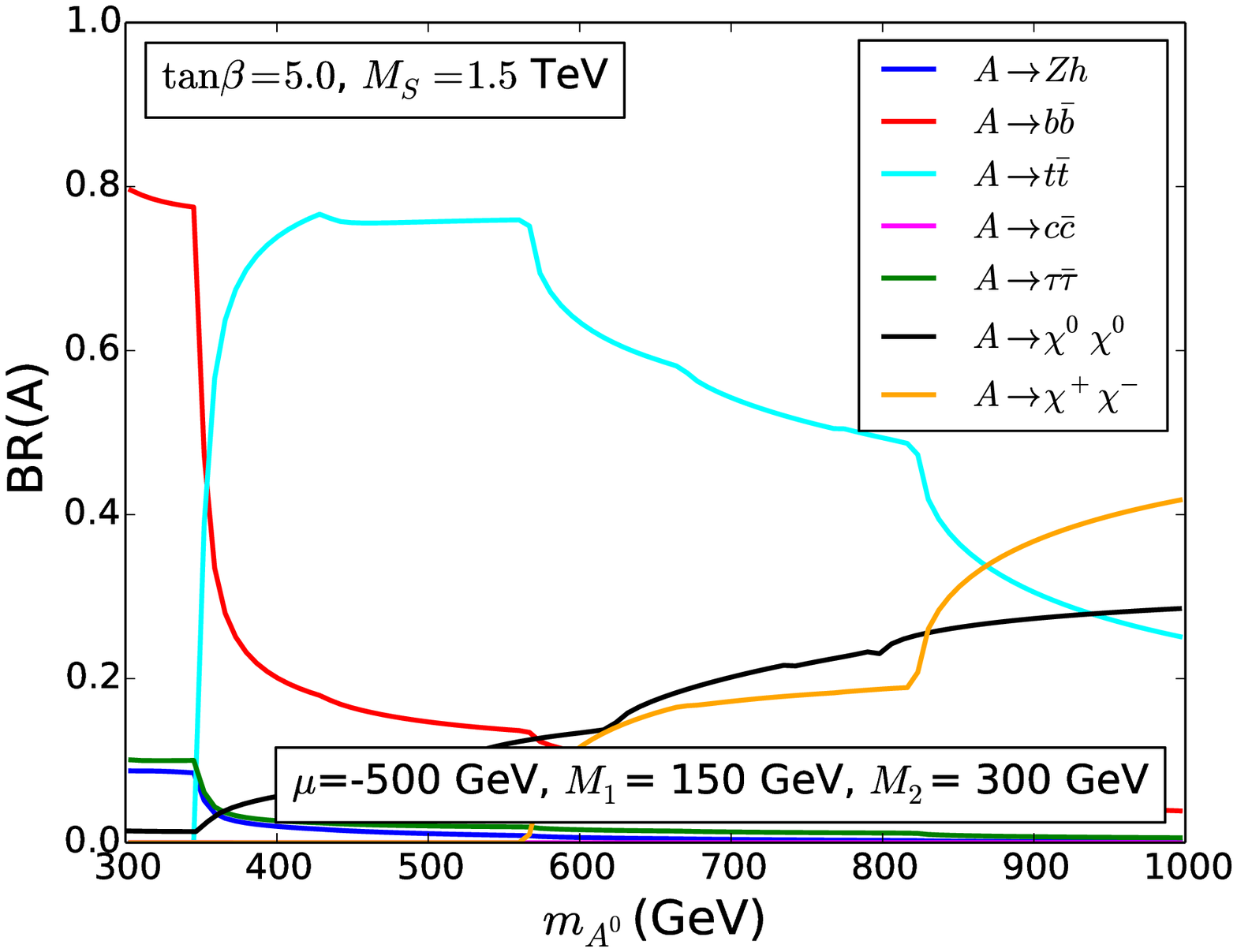}}
\hspace{-.8cm}\centering \subfigure[]{
\includegraphics[scale=0.41]{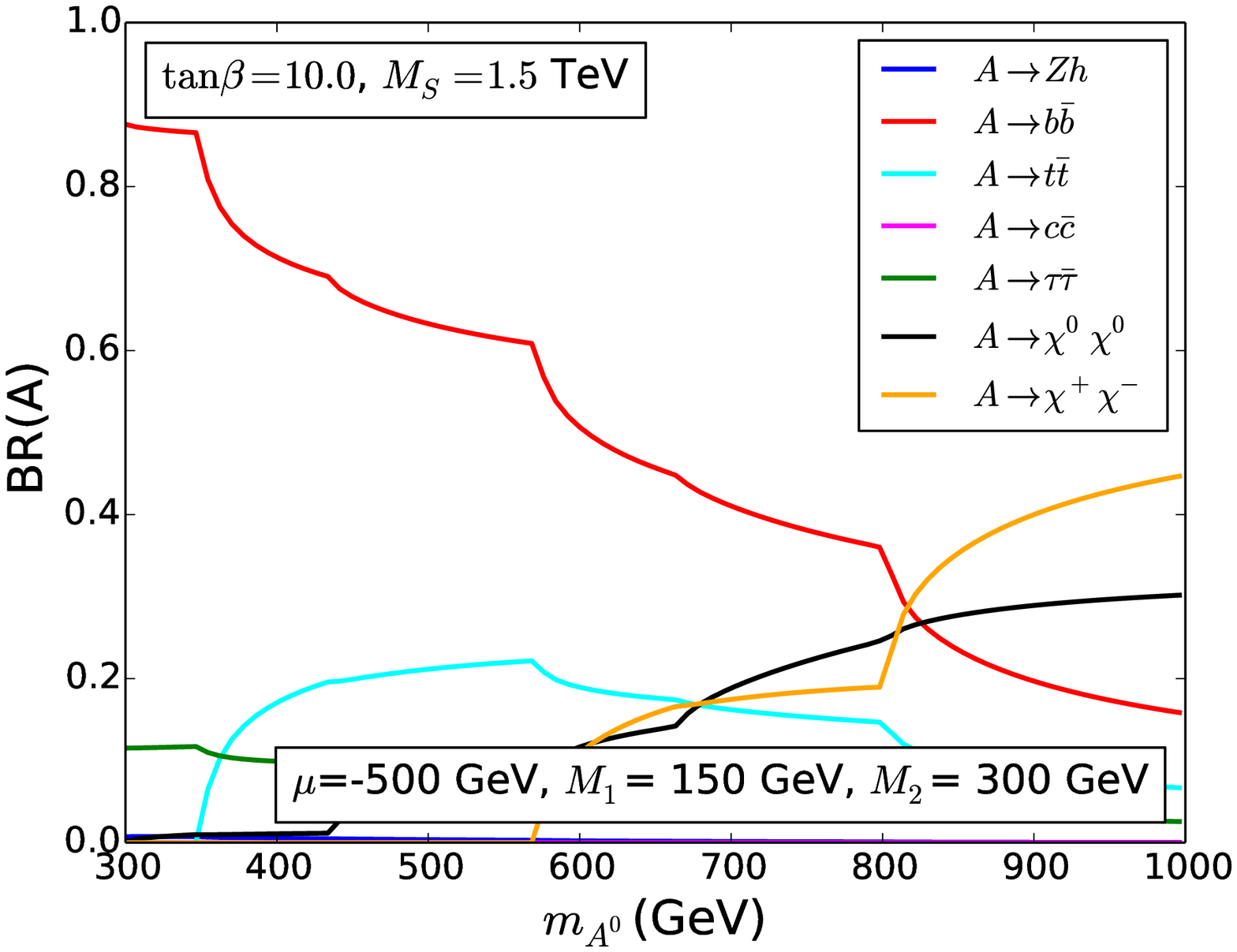}}
\caption{Branching fractions for different decay channels of $A$  as the
 functions of $m_{A^0}$ for the fixed values of parameters
 shown in the inset.}\label{adecay}
\end{figure}

\begin{figure}[t]
\centering \subfigure[]{
\includegraphics[scale=0.34]{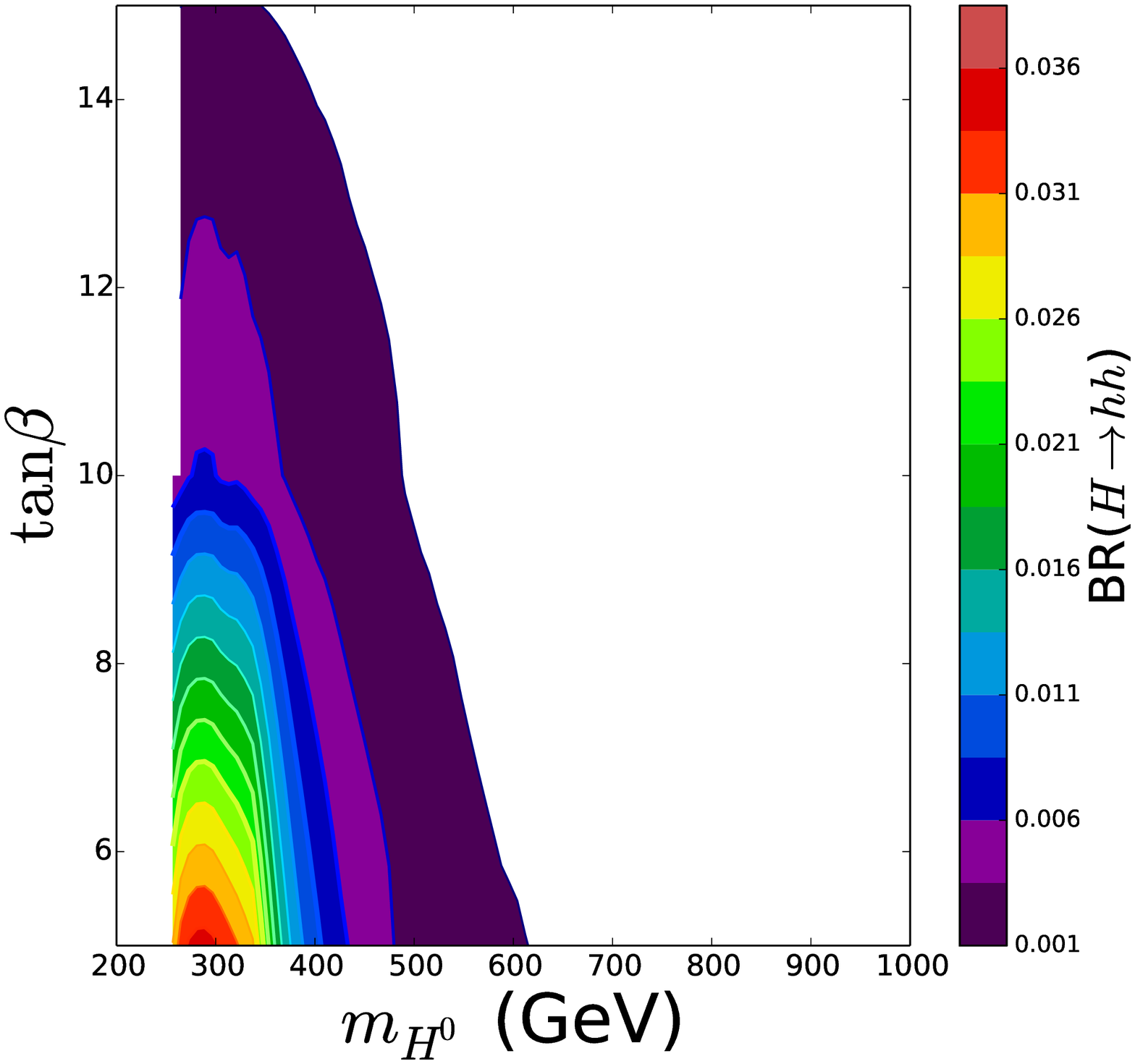}}
\hspace{-.8cm}\centering \subfigure[]{
\includegraphics[scale=0.34]{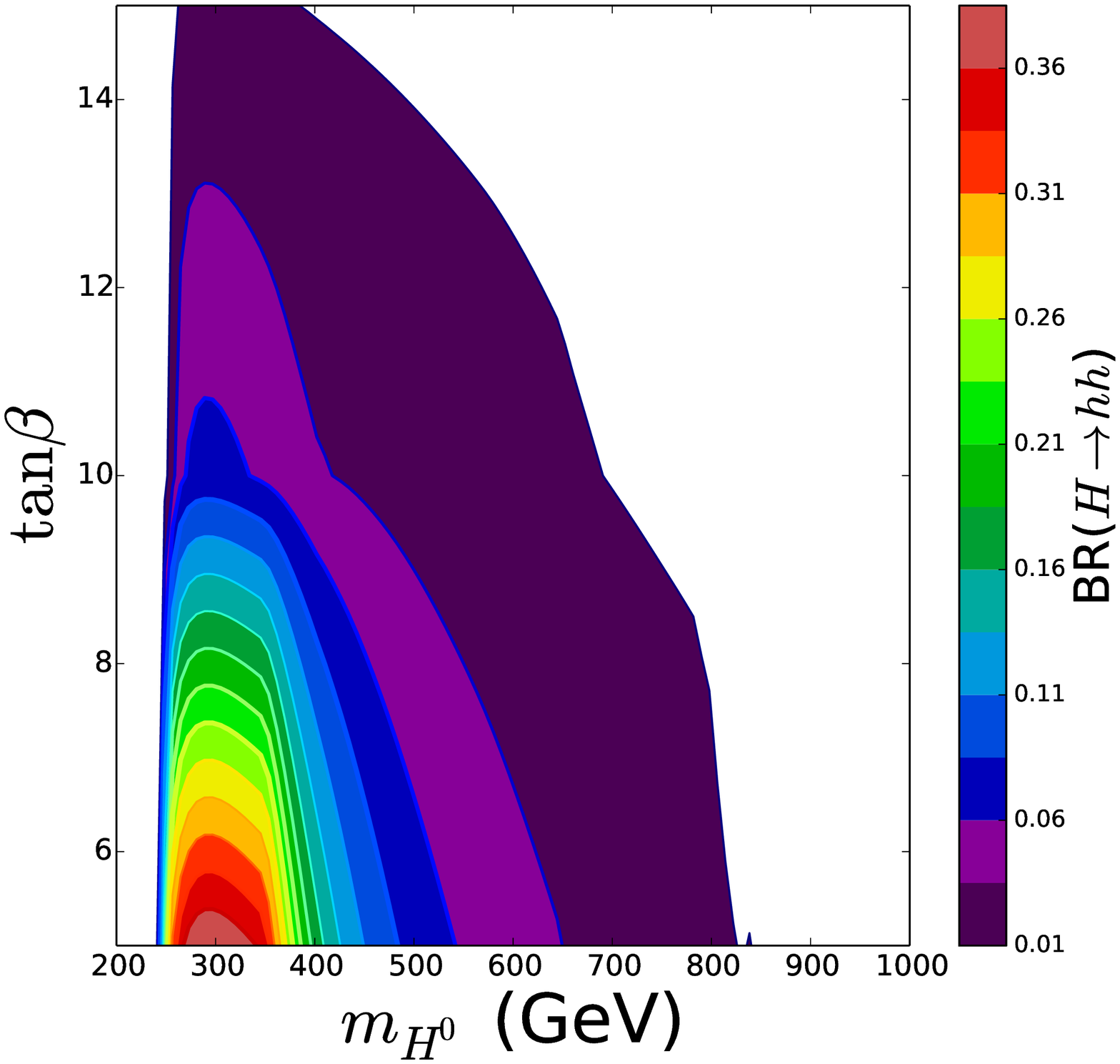}}
\caption{The contours of constant branching ratio for $H$ decay to  $hh$
 for (a) $\mu$ = -230 GeV, $M_1$ = 120 GeV and $M_2$ = 240 GeV, and
 (b) $\mu$ = -500 GeV, $M_1$ = 150 GeV and $M_2$ = 300 GeV. Other
 parameters are varied  to get the
 lightest Higgs mass in 122-128 GeV range.}\label{conthdecay}
\end{figure}

In Figs. \ref{hdecay} and \ref{adecay}, we
 show the branching fractions of heavy Higgs and pseudoscalar to different
 channels respectively for $\tan\beta$ = 5, 10. In this paper we have also
 considered  supersymmetric particles in the final states, consistent with
 current experimental constraints. All the SUSY particles have masses $\approx$ $M_S$ except
 neutralinos and charginos. The neutralinos and charginos mass spectrum depends on
 the supersymmetry breaking gaugino mass parameters $M_1$ and $M_2$ as well
 as $\mu$ and $\tan\beta$. We recall that the chargino mass lower bounds from
 the LEP experiment imply \cite{leplimit}
\be  |\mu|, M_2 \ge 100 \,\, \mbox{GeV}. \ee
We shall confine ourselves
 to the scenario where supersymmetry breaking gaugino
 masses $M_i$ ($i$ = 1, 2, 3) are equal at the grand unified scale. 
In this case the renormalization group evolution of $M_i$ implies
 $M_1$ $\approx$ 0.5 $M_2$ at the weak scale. We shall consider the
 parameter space consistent with these constraints.

 We have plotted the $H$ and $A$ branching fractions for
 two benchmark values of $\mu$, $M_1$ and $M_2$. The neutralino and chargino
 mass spectrum for the  benchmark points is given in Table \ref{table1}.
The second benchmark point has comparatively heavy spectrum.
We observe that for $m_{A^0}$ $\approx$ $m_{H^0}$ $\geq$ 500 GeV,
 neutralino and chargino are the dominant decay modes of both
 $H$ and $A$, for a light neutralino and chargino spectrum. If these
 are heavy then $t \bar t$ is the dominant decay channel for
 low values of $\tan\beta$.  In case of $H$ decay, below $t \bar{t}$
 threshold both $b \bar{b}$ and $hh$ have appreciable branching fractions. For
 large $\tan\beta$ and heavy neutralino and chargino
 spectrum, $b \bar{b}$ is the dominant decay mode of the heavy
 Higgs boson as shown in Fig. \ref{hdecay} (d). Our 
 aim is to study BR($H$ $\rightarrow$ $hh$) and BR($A$ $\rightarrow$ $hZ$), since the former
 involves the Higgs trilinear coupling and latter is background for
 multiple Higgs production processes. We can see from Fig. \ref{adecay} that
 $A \rightarrow h Z$ branching fraction is negligible for values of  $\tan\beta$ = 5
 and 10. Below $\chi^+\chi^-$ threshold, $A$ $\rightarrow$ $b \bar{b}$ is
 the dominant decay channel for large value of $\tan\beta$.

 The contours of constant values of BR($H$ $\rightarrow$ $hh$ )
 are shown in Fig. \ref{conthdecay} for the benchmark points. The
 BR($H$ $\rightarrow$ $hh$) decreases with increasing $\tan\beta$. Since the
 neutralino and chargino spectrum is heavy for the second benchmark point,
 BR($H \rightarrow \chi^0 \chi^0, \chi^+ \chi^-$) is
 suppressed as compared to first benchmark point and
 consequently BR($H \rightarrow hh$) is enhanced.  In all the plots
 we have varied the relevant parameters in a manner so that the
 lightest Higgs mass is in the range 122-128 GeV. The
 main parameter adjusted in this context is $A_t$. As already mentioned
 we have allowed a 3 GeV theoretical uncertainty in the Higgs mass 
calculations. If one wants to restrict to the range of 124-126 GeV for the 
mass of the Higgs boson, then we will have a corresponding slightly narrow 
band of $A_t$ values. In other words that will also reduce the range
 of values of the parameter $A_t$. But that minor variation in the $A_t$ values will not 
change our analysis significantly since $A_t$ parameter enters through 
one loop radiative corrections in the trilinear coupling calculations. Processes
 shown in Fig. \ref{feyn1} involve only
 trilinear coupling $\lambda_{Hhh}$, but Fig. \ref{feyn2} (c) and
 \ref{feyn3} (c) involve both $\lambda_{hhh}$ and $\lambda_{Hhh}$. Therefore
 one has to study non-resonant multiple Higgs production
 cross-section to measure $\lambda_{hhh}$ coupling.

\section{Measurement of Trilinear Higgs Couplings \label{tricoupmeasurement}}
\begin{figure}[t]
\centering \subfigure[]{
\includegraphics[scale=0.34]{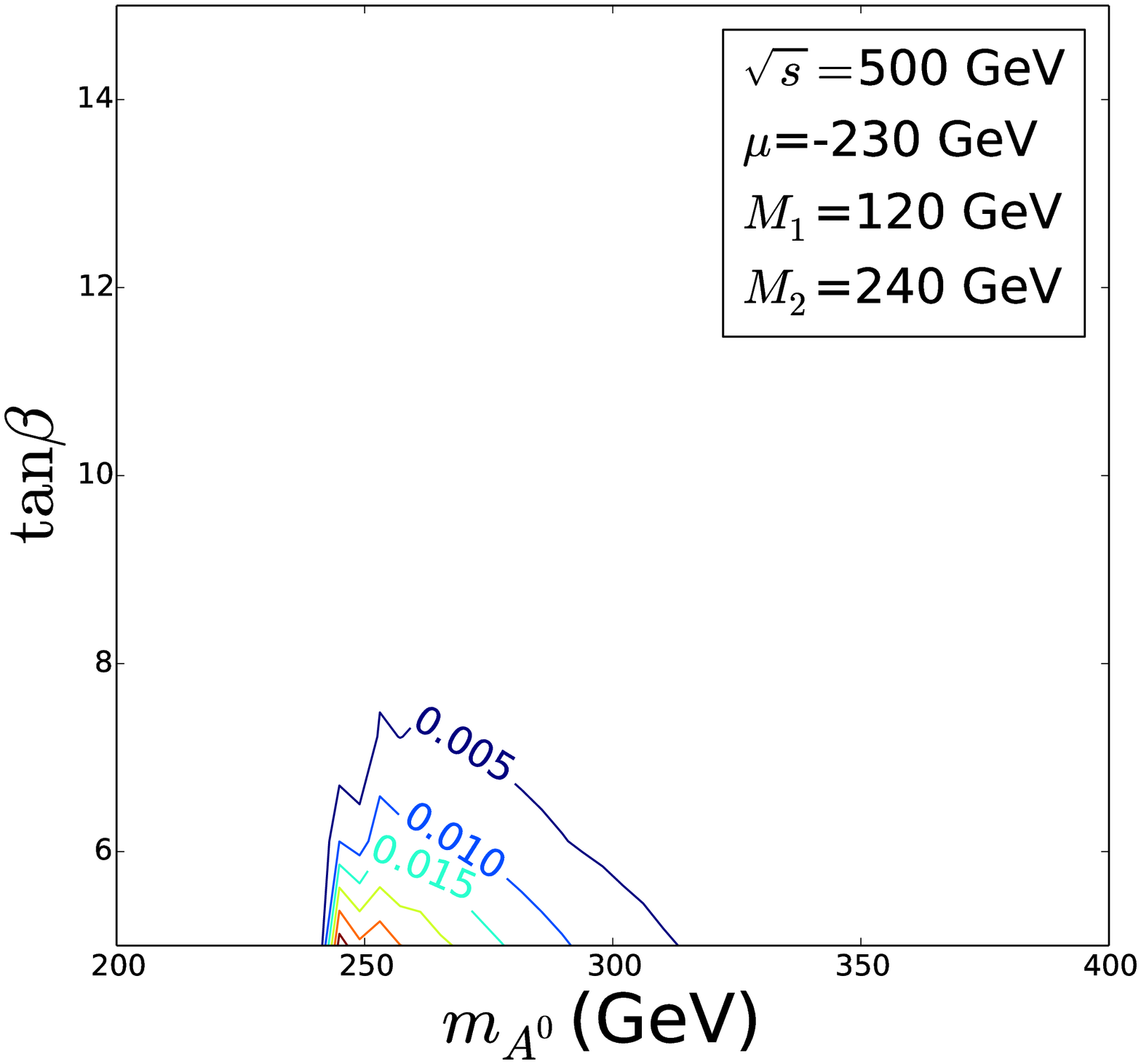}}
\hspace{-.8cm}\centering \subfigure[]{
\includegraphics[scale=0.34]{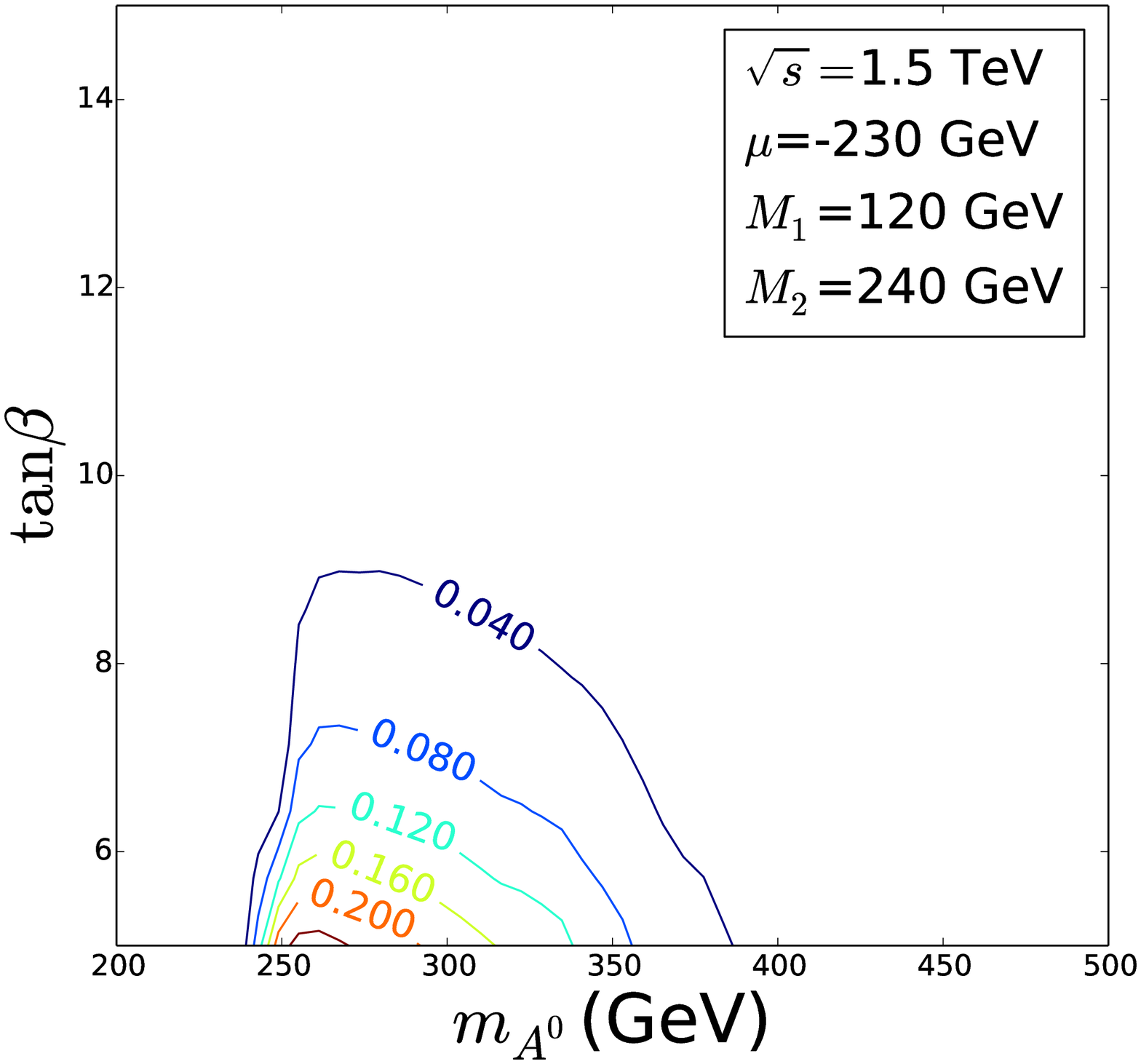}}
\centering \subfigure[]{
\includegraphics[scale=0.34]{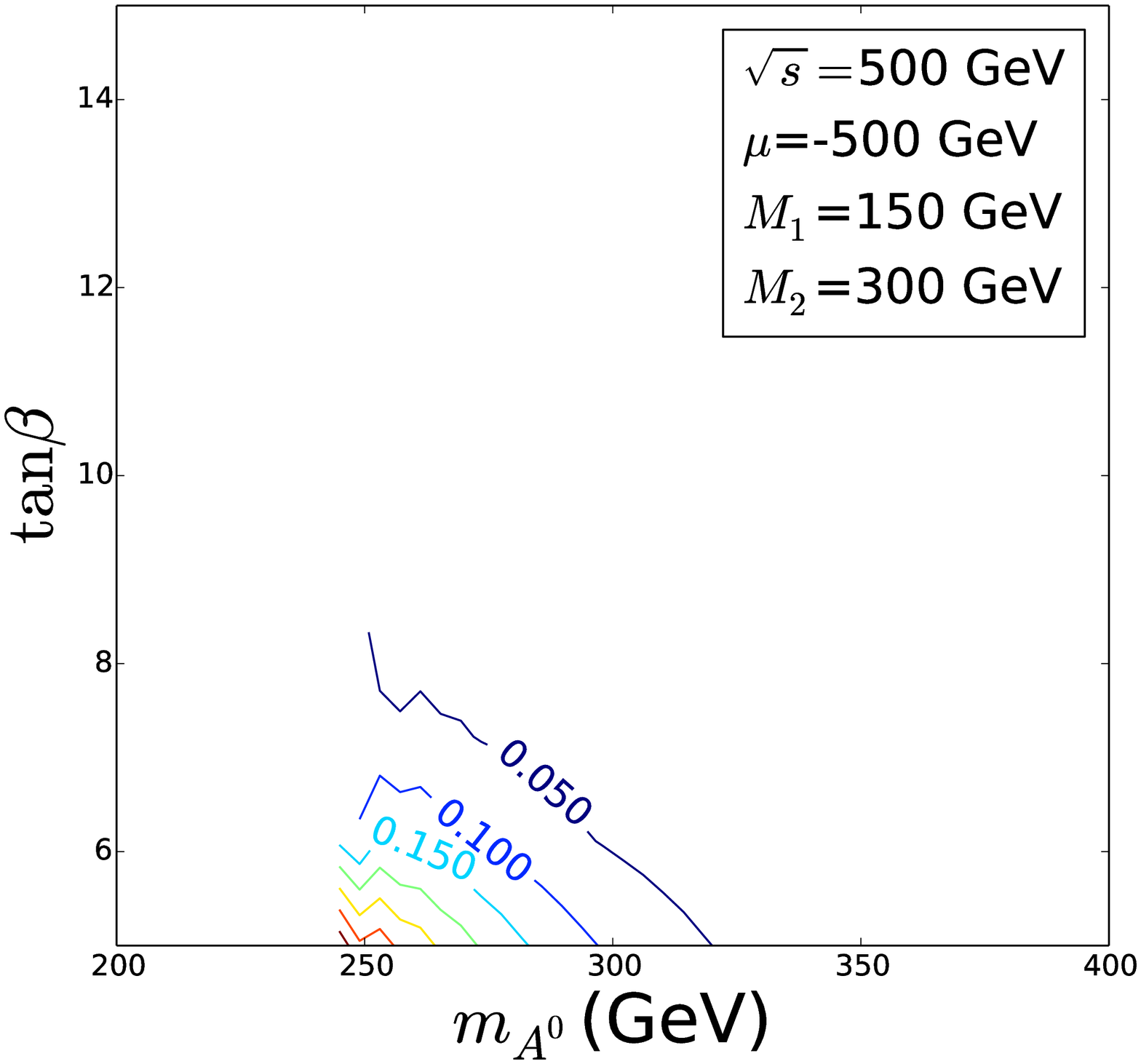}}
\hspace{-.8cm}\centering \subfigure[]{
\includegraphics[scale=0.34]{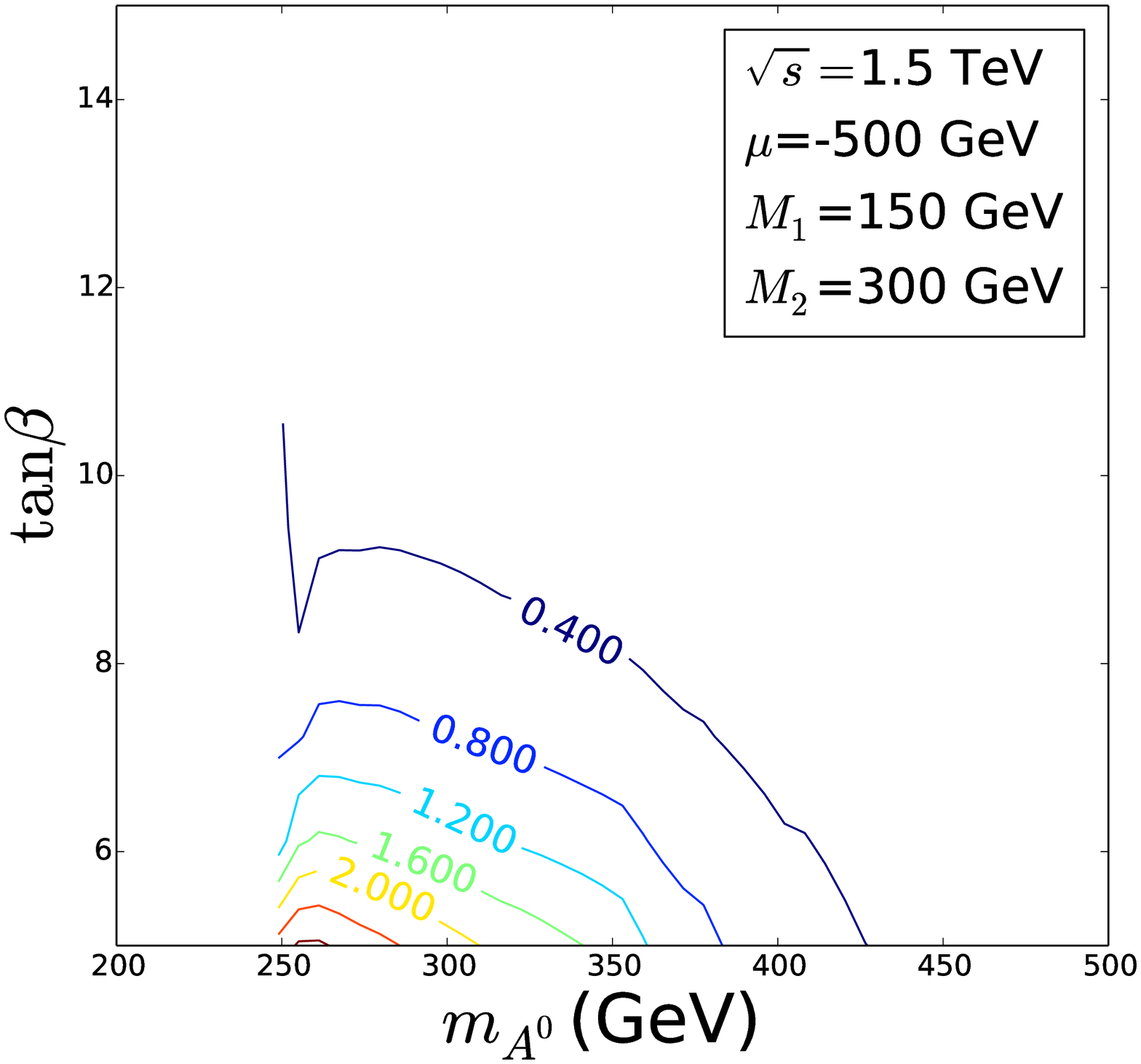}}
\caption{The contours of constant $\sigma(H) \times BR(H \rightarrow hh)$ (in fb)
 for the values of the parameters shown in the inset.}\label{tcregion}
\end{figure}

In this Section we will compute the regions of ($m_{A^0}$,$\tan\beta$) plane
 where trilinear couplings $\lambda_{Hhh}$ and $\lambda_{hhh}$ can be measured.
We calculate the heavy Higgs production cross-section using
 Eqs. \ref{Eq:sigZH}, \ref{Eq:sigAH} and \ref{Eq:fusion-exact}. The
 contours of $\sigma(H) \times BR(H \rightarrow hh)$ for $\sqrt{s}$ = 500 GeV
 and $\sqrt{s}$ = 1.5 TeV, respectively are shown in Fig. \ref{tcregion}. In
 upper left and right panel, the outermost contours correspond to
 $\sigma(H) \times BR(H \rightarrow hh )$ $\approx$ 0.005 fb and .04 fb,
 respectively. The $\sigma(H) \times BR(H \rightarrow hh )$ decreases as we
 move diagonally upward in the $(m_{A^0}$,$\tan\beta)$ plane because
 $BR(H \rightarrow hh)$ decreases in this direction. As shown in
 Fig. \ref{crossplots}, the heavy Higgs production cross-section increases for
 $\sqrt{s}$ = 1.5 TeV, therefore one can measure $\lambda_{Hhh}$ coupling with a lower luminosity.

The BR($H \rightarrow hh$) is directly proportional to $\lambda_{Hhh}$ and
 this branching ratio decreases with increasing $\tan\beta$ for fixed value of
 $m_{A^0}$. As discussed earlier, heavy Higgs production cross-section
decreases with $m_{A^0}$. The $H$ branching ratio to $hh$ pair
 is kinematically forbidden for $m_{A^0}$ $\approx$ $m_{H^0}$ $\le$ 250
 GeV. Therefore, the lower left corner of $(m_{A^0}$,$\tan\beta)$ plane is the
 suitable region to measure $\lambda_{Hhh}$ coupling. We can see from
 the Fig. \ref{tcregion} that $\sigma(H) \times BR(H \rightarrow hh )$
 is sensitive to the $\mu$ parameter. In other words, precise
 knowledge of neutralino and chargino spectrum is crucial in order to determine the
 $\lambda_{Hhh}$ coupling. For 
 $\sqrt{s}$ = 1.5 TeV and heavy neutralino and chargino spectrum with
 $\tan\beta$ $\approx$  10, $\sigma(H) \times BR(H \rightarrow hh )$ $\approx$ 0.4 fb. This
 cross-section is 0.05 fb for $\sqrt{s}$=500 GeV so with a luminosity of 500 fb $^{-1}$, 25 events
 could be seen. This indicates only order of
 magnitude that can be reached but the actual number of events seen will be
 lowered by the efficiencies. The simulations of signal and background will depend on the detector
 sensitivity which is not the focus of this paper.

\begin{figure}[t]
\centering \subfigure[]{
\includegraphics[scale=0.34]{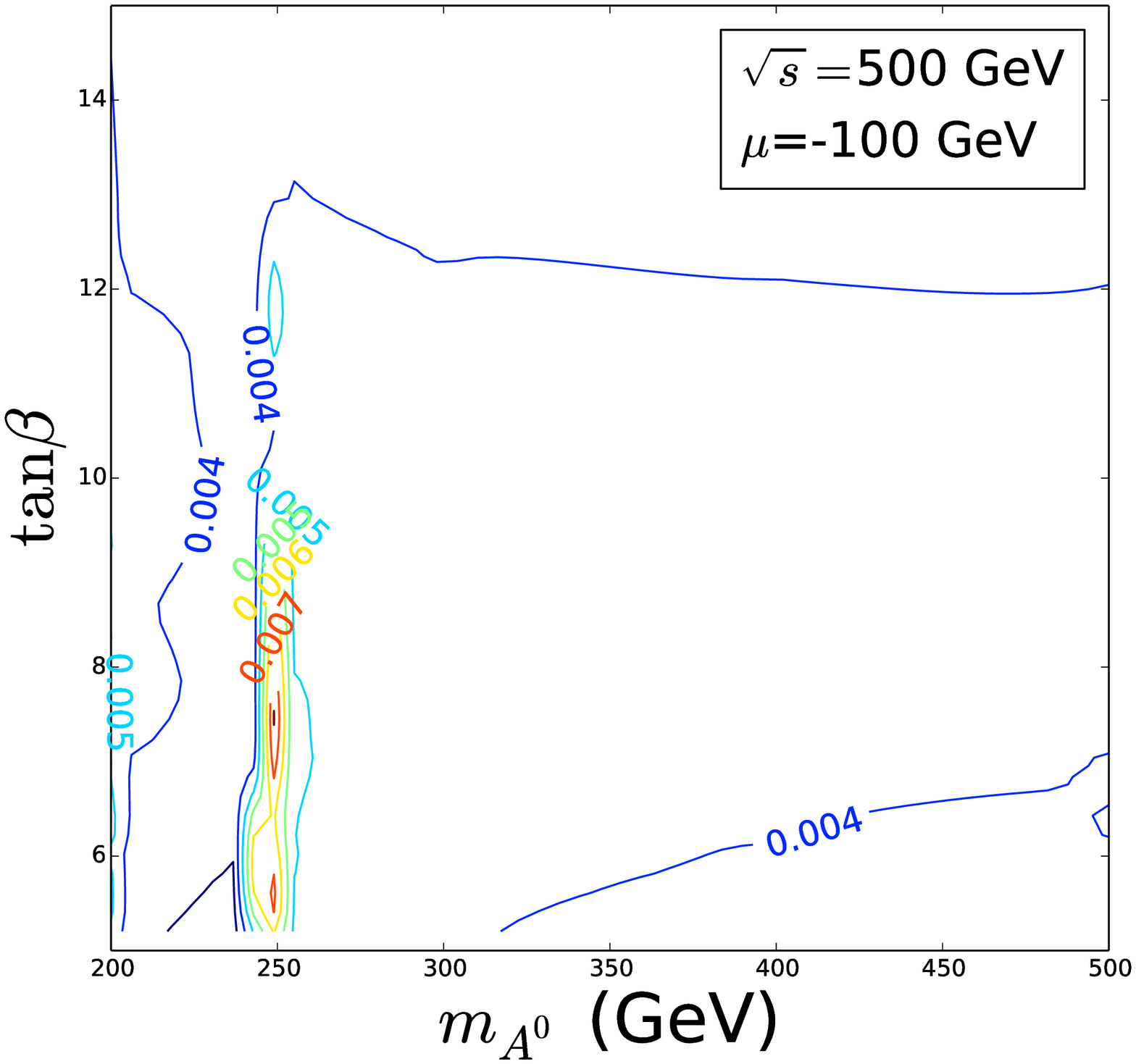}} 
\hspace{-.8cm}\centering \subfigure[]{
\includegraphics[scale=0.34]{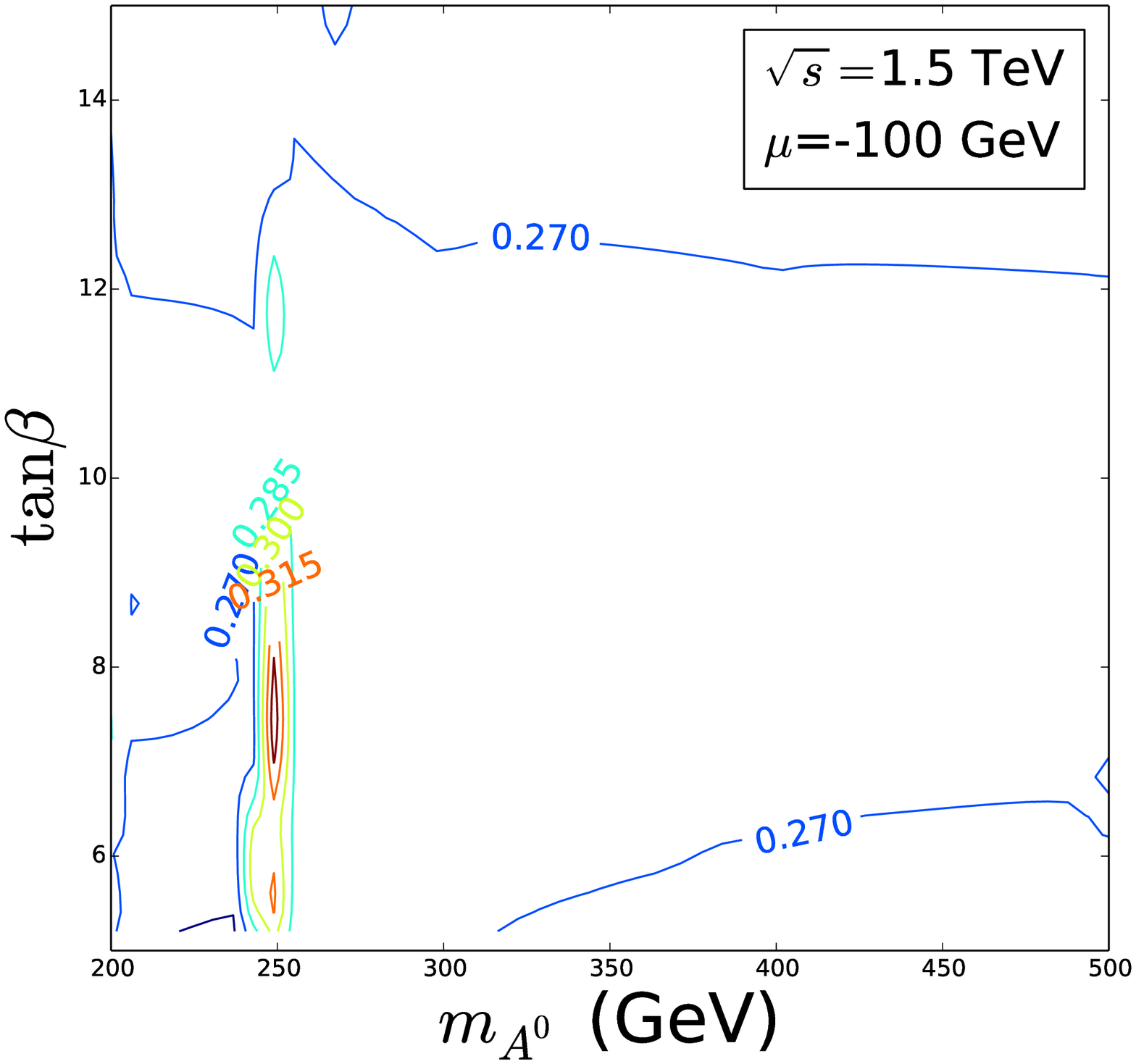}}
\caption{The contours of non-resonant $\sigma(ee \rightarrow hh \nu \bar\nu) $ (in fb)
 via non-resonant WW fusion.}\label{tcregionhhh}
\end{figure}

The $hh \nu_e \bar\nu_e$ final state produced through non-resonant $WW$ fusion
 involves both $\lambda_{hhh}$ and $\lambda_{Hhh}$ coupling. Having an estimate
 of the coupling $\lambda_{Hhh}$ from resonant $WW$ production process, we
 can use non-resonant $WW$ fusion process to measure the trilinear $\lambda_{hhh}$ coupling.
 In Fig. \ref{tcregionhhh} we show  the constant contours of the cross-section
 for the non-resonant $WW$ fusion  process $\sigma(hh\nu_e\bar\nu_e)$ 
 in the $(m_{A^0},\tan\beta)$ plane.  We can see that the
  cross-section $\sigma(hh\nu_e \bar\nu_e)$ is
 almost independent of the values of $m_{A^0}$ and $\tan\beta$. The chances
 of the measurement of $\lambda_{hhh}$ are same in most of the
  $(m_{A^0},\tan\beta)$ parameter
 space. There is a small increase in the cross-section at the
 boundary $m_{H^0}^2$ $\approx$ 4 $m_{h^0}^2$ where Fig. \ref{feyn2}(c) starts
 contributing through
 resonant process. Even with the 1000 fb $^{-1}$ of  luminosity
 at $\sqrt{s}$ = 500 GeV one could see only few events.

In Fig. \ref{crossplots} we have plotted the cross-section for the background process
 $ e^+ e^- \rightarrow A h$  and BR ($A \rightarrow hZ$) in Fig. \ref{adecay}. Since
 BR ($A \rightarrow hZ$) is negligible this process will be suppressed for the
 considered parameter space. Also this kind of background events
 can be easily distinguished from the signal events by just looking
 at the $hh$ pair invariant mass distribution which will resonate
 in case of signal process.

\section{Conclusions\label{Conclusions}}
We have carried out a detailed analysis of the measurement of trilinear
 couplings of the neutral CP-even Higgs boson, $\lambda_{Hhh}$ and $\lambda_{hhh}$,
 at an electron-positron collider. For this purpose we have identified
 the state observed at CERN Large Hadron Collider at $\approx$ 125 GeV with
 the lightest Higgs boson of the MSSM. This identification has been used to study 
the dependence of the mass of the heavier CP-even Higgs boson of the MSSM ($H^0$) 
on the parameter space of the MSSM, so as to get a handle on the mass of $H^0$. 
Furthermore, we have also used the lower bound on the chargino mass from
 the LEP experiments to constrain the parameter space of the MSSM. All these
 constraints have been used in our study of the trilinear couplings.

Our main purpose is to investigate various processes involving multiple Higgs
 bosons in the final state in $e^+$-$e^-$ collisions, consistent with the
 constraints summarized above. The production of the
 heavier Higgs bosons in $e^+ e^-$ collisions can lead to multiple lighter
 Higgs bosons ($h^0$) in the final state, which can be used in the measurement
 of the trilinear couplings of the CP-even Higgs bosons.

 We indicate the regions of the $(m_{A^0},\tan\beta)$ plane where trilinear
 coupling $\lambda_{Hhh}$ and $\lambda_{hhh}$ can be measured
 at the linear collider. The resonant heavy Higgs production processes are
 used to extract $\lambda_{Hhh}$ coupling. For $\sqrt{s}$= 1.5 TeV, $\tan\beta$ $\approx$ 8, 
the $\sigma(H) \times BR (H \rightarrow hh )$ $\approx$ 1 fb, and
  regions of $\tan\beta$ upto 10 and $m_{A^0}$ upto 450
 can be explored for $\lambda_{Hhh}$ coupling measurement. However high
 luminosity is required to probe larger $\tan\beta$ values. For the measurement
 of the $\lambda_{hhh}$ coupling, we use light Higgs pair production
 through non-resonant WW fusion, and this cross-section is not very
 sensitive to $m_{A^0}$ and $\tan\beta$. Besides values
 of $m_{A^0}$ and $\tan\beta$, the information
 of neutralino and chargino masses is crucial for
 determining the trilinear couplings.  
\section{Acknowledgments}
The work of P. N. Pandita is supported by the Department of  Atomic Energy, India
through its Raja Ramanna Fellowship. He would like to thank the 
Inter University Centre for Astronomy and Astrophysics, Pune for hospitality 
where part of this work was done.  C. K. Khosa would like to thank  Jayita
 Lahiri for many fruitful discussions.
\section{Appendix \label{appendixtc}}
In this Appendix we summarize the one-loop radiative corrections to
 the trilinear Higgs couplings in the MSSM \cite{OPtricooup,Osland:1999ae,Osland:1999qw,Barger:1991ed}. 
The radiative corrections, in units
 of $(\sqrt{2} G_F)^{1/2} {m_Z^2}$ can be written as 
\begin{eqnarray}
\Delta \lambda_{Hhh} & = & \left( \frac{3g^2 \cos^2\theta_W}{16
\pi^2} \frac{m_t^4}{m_W^4}
\frac{\sin\alpha\cos^2\alpha}{\sin^3\beta} \right)
\nonumber\\
& \times & \left[ 3\log\frac{m_{\tilde t_1}^2 m_{\tilde
t_2}^2}{m_t^4} + (m_{\tilde t_1}^2 - m_{\tilde t_2}^2)C_t(E_t +
2F_t)
   \log\frac{m_{\tilde t_1}^2}{m_{\tilde t_2}^2} \right. \nonumber\\
& & + 2\left(\frac{m_t^2}{m_{\tilde t_1}^2} \left[1+(m_{\tilde
t_1}^2 - m_{\tilde t_2}^2)C_t E_t\right] \left[1+(m_{\tilde t_1}^2
- m_{\tilde t_2}^2)C_t F_t\right]^2 \right.
\nonumber \\
& & + \left. \left.\frac{m_t^2}{m_{\tilde t_2}^2}
\left[1-(m_{\tilde t_1}^2 - m_{\tilde t_2}^2)C_t E_t\right]
\left[1-(m_{\tilde t_1}^2 - m_{\tilde t_2}^2)C_t F_t\right]^2 - 2
\right)
\right], \label{17}  \\
\Delta \lambda_{hhh} & = & \left( \frac{3g^2 \cos^2\theta_W}{16
\pi^2} \frac{m_t^4}{m_W^4} \frac{\cos^3\alpha}{\sin^3\beta}
\right)
\nonumber \\
& \times & \left[ 3\log\frac{m_{\tilde t_1}^2 m_{\tilde
t_2}^2}{m_t^4} + 3 (m_{\tilde t_1}^2 - m_{\tilde t_2}^2)C_t F_t
   \log\frac{m_{\tilde t_1}^2}{m_{\tilde t_2}^2} \right. \nonumber\\
& & + \left. 2\left(\frac{m_t^2}{m_{\tilde t_1}^2}
\left[1+(m_{\tilde t_1}^2 - m_{\tilde t_2}^2)C_t F_t\right]^3
+\frac{m_t^2}{m_{\tilde t_2}^2} \left[1-(m_{\tilde t_1}^2 -
m_{\tilde t_2}^2)C_t F_t\right]^3
-2 \right) \right], \label{16} \\  \\
\Delta\lambda_{hAA} & = & \left( \frac{3g^2 \cos^2\theta_W}{16
\pi^2} \frac{m_t^4}{m_W^4} \frac{\cos\alpha
\cos^2\beta}{\sin^3\beta} \right)
\nonumber\\
& \times & \biggl[ \log\frac{m_{\tilde t_1}^2 m_{\tilde
t_2}^2}{m_t^4} +  (m_{\tilde t_1}^2 - m_{\tilde t_2}^2)(D_t^2 +
C_t F_t)
   \log\frac{m_{\tilde t_1}^2}{m_{\tilde t_2}^2}  \nonumber\\
& & + (m_{\tilde t_1}^2 - m_{\tilde t_2}^2)^2 C_tD_t^2F_t\,
g(m_{\tilde t_1}^2,m_{\tilde t_2}^2) \biggr],      \label{21} \\
\Delta \lambda_{HAA} & = & \left( \frac{3g^2 \cos^2\theta_W}{16
\pi^2} \frac{m_t^4}{m_W^4}
\frac{\sin\alpha\cos^2\beta}{\sin^3\beta} \right)
\nonumber\\
& \times & \left[\log\frac{m_{\tilde t_1}^2 m_{\tilde
t_2}^2}{m_t^4} + (m_{\tilde t_1}^2 - m_{\tilde t_2}^2)(D_t^2 +
C_t E_t) \log\frac{m_{\tilde t_1}^2}{m_{\tilde t_2}^2} \right. \nonumber\\
& &\left. +  (m_{\tilde t_1}^2 - m_{\tilde t_2}^2)^2 C_t D_t^2 E_t
g(m_{\tilde t_1}^2,m_{\tilde t_2}^2) \right], \label{22}
\end{eqnarray}
\begin{eqnarray}
\Delta \lambda_{HHH} & = & \left( \frac{3g^2 \cos^2\theta_W}{16
\pi^2} \frac{m_t^4}{m_W^4} \frac{\sin^3\alpha}{\sin^3\beta}
\right)
\nonumber\\
& \times & \left[ 3\log\frac{m_{\tilde t_1}^2 m_{\tilde
t_2}^2}{m_t^4} +3(m_{\tilde t_1}^2 - m_{\tilde t_2}^2)C_t E_t
   \log\frac{m_{\tilde t_1}^2}{m_{\tilde t_2}^2} \right. \nonumber\\
& & + 2\left(\frac{m_t^2}{m_{\tilde t_1}^2} \left[1+(m_{\tilde
t_1}^2 - m_{\tilde t_2}^2)C_t E_t\right]^3  \right.
\nonumber \\
& & + \left. \left.\frac{m_t^2}{m_{\tilde t_2}^2}
\left[1-(m_{\tilde t_1}^2 - m_{\tilde t_2}^2)C_t E_t\right]^3 - 2
\right) \right], \label{23} \\
\Delta \lambda_{HHh} & = & \left( \frac{3g^2 \cos^2\theta_W}{16
\pi^2} \frac{m_t^4}{m_W^4}
\frac{\sin^2\alpha\cos\alpha}{\sin^3\beta} \right)
\nonumber\\
& \times & \left[ 3\log\frac{m_{\tilde t_1}^2 m_{\tilde
t_2}^2}{m_t^4} + (m_{\tilde t_1}^2 - m_{\tilde t_2}^2)C_t(2 E_t +
F_t)
   \log\frac{m_{\tilde t_1}^2}{m_{\tilde t_2}^2} \right. \nonumber\\
& & + 2\left(\frac{m_t^2}{m_{\tilde t_1}^2} \left[1+(m_{\tilde
t_1}^2 - m_{\tilde t_2}^2)C_t E_t\right]^2 \left[1+(m_{\tilde
t_1}^2 - m_{\tilde t_2}^2)C_t F_t\right] \right.
\nonumber \\
& & + \left. \left.\frac{m_t^2}{m_{\tilde t_2}^2}
\left[1-(m_{\tilde t_1}^2 - m_{\tilde t_2}^2)C_t E_t\right]^2
\left[1-(m_{\tilde t_1}^2 - m_{\tilde t_2}^2)C_t F_t\right] - 2
\right) \right], \label{24}
\end{eqnarray}
where
\begin{eqnarray}
C_t & = & (A + \mu\cot\beta)/(m_{\tilde t_1}^2 - m_{\tilde
t_2}^2),
\nonumber\\
D_t & = & (A - \mu\tan\beta)/(m_{\tilde t_1}^2 - m_{\tilde
t_2}^2),
\nonumber\\
E_t & = & (A + \mu\cot\alpha)/(m_{\tilde t_1}^2 - m_{\tilde
t_2}^2),
\nonumber\\
F_t & = & (A - \mu\tan\alpha)/(m_{\tilde t_1}^2 - m_{\tilde
t_2}^2), \label{Eq:CEF}
\end{eqnarray}

\begin{equation}
g(m_{\tilde t_1}^2, m_{\tilde t_2}^2) =  2 - \frac{m_{\tilde
t_1}^2 + m_{\tilde t_2}^2}{m_{\tilde t_1}^2 - m_{\tilde t_2}^2}
\log\frac{m_{\tilde t_1}^2}{m_{\tilde t_2}^2}. \label{7}
\end{equation}

\end{document}